\documentclass[11pt,a4paper]{article}

\usepackage{amsmath,amsthm,latexsym,amssymb}
\usepackage{graphicx,bm,fancybox,multirow,hhline,indentfirst}
\usepackage{ascmac,atbegshi}
\usepackage{natbib,verbatim,color}
\usepackage{booktabs,subfig,url}
\usepackage[colorlinks,citecolor=blue,urlcolor=blue]{hyperref}

\makeatletter
\def\section{\@startsection {section}{1}{\z@}{-3.5ex plus -1ex minus -.2ex}{2.3 ex plus .2ex}{\large\sc\centering}}
\def\subsection{\@startsection {subsection}{1}{\z@}{-3.5ex plus -1ex minus -.2ex}{2.3 ex plus .2ex}{\large}}
\makeatother

\theoremstyle{definition}
\newtheorem{theorem}{Theorem}
\newtheorem{lemma}{Lemma}
\newtheorem*{remark}{Remark}
\newcommand{\argmin}{\mathop{\rm argmin}}
\newcommand{\indep}{\mathop{\perp\!\!\!\!\perp}}

\title{\Large\bf Selective Inference in Propensity Score Analysis \medskip}
\author{Yoshiyuki Ninomiya\\{Department of Statistical Inference and Mathematics}\\{The Institute of Statistical Mathematics}
\medskip \and Yuta Umezu\\{School of Information and Data Science, Nagasaki University} 
\medskip \and Ichiro Takeuchi\\{Department of Computer Science, Nagoya Institute of Technology}\\{RIKEN Center for Advanced Intelligence Project} 
}
\date{}

\selectfont

\allowdisplaybreaks

\def\E{{\rm E}}
\def\P{{\rm P}}
\def\V{{\rm V}}
\def\oP{{\rm o}_{\rm P}}
\def\OP{{\rm O}_{\rm P}}

\begin{document}

\maketitle

\begin{abstract}
Selective inference (post-selection inference) is a methodology that has attracted much attention in recent years in the fields of statistics and machine learning. Naive inference based on data that are also used for model selection tends to show an overestimation, and so the selective inference conditions the event that the model was selected. In this paper, we develop selective inference in propensity score analysis with a semiparametric approach, which has become a standard tool in causal inference. Specifically, for the most basic causal inference model in which the causal effect can be written as a linear sum of confounding variables, we conduct Lasso-type variable selection by adding an $\ell_1$ penalty term to the loss function that gives a semiparametric estimator. Confidence intervals are then given for the coefficients of the selected confounding variables, conditional on the event of variable selection, with asymptotic guarantees. An important property of this method is that it does not require modeling of nonparametric regression functions for the outcome variables, as is usually the case with semiparametric propensity score analysis.

\

\end{abstract}

\newpage

\section{Introduction}
\label{sec1}
It has often been pointed out that generating a hypothesis of interest from data and then using the same data to construct a statistical test or a confidence interval for that hypothesis is inappropriate. \cite{Bre01} called it a ``quiet scandal of statistics.'' The so-called ``file drawer problem'' can be regarded as this kind of problem. A typical example in regression is naive inference on the coefficients of the selected variables after variable selection. To deal with this variable selection problem, ``simultaneous inference'' was proposed by \cite{BerBBZZ13}. Simultaneous inference gives a reasonable inference no matter what model is selected, but as one can imagine from the methods of multiple testing, it is rather conservative. In these situations, a ``selective inference'' approach, which is based on the work of \cite{LeeT14}, \cite{LeeSST16}, and \cite{TibTLT16}, has recently gained attention in the field of statistics and machine learning. Selective inference conditions on that the model has been selected. At about the same time as these papers were published, \cite{TayT15} explained the great potential of selective inference. It has been extensively developed since then.

On the other hand, causal inference has been a hot topic in biostatistics and econometrics and has recently been of importance in machine learning (see, e.g., \citealt{PetJS17}; \citealt{HerR20}). In particular, propensity score analysis with a semiparametric approach has become a standard tool, as it is a theoretically guaranteed method that avoids the bias introduced by the existence of confounding. To illustrate the motivation for this paper, let us examine one of the simplest models in causal inference, where the causal effect can be written as a linear sum of confounding variables, as follows:
\begin{align*}
Y_i=T_i\bm{X}_i'\bm{\beta}^{(1)}+(1-T_i)\bm{X}_i'\bm{\beta}^{(0)}+\xi_i
\qquad (1\le i\le n).
\end{align*}
The subscript $i\ (\in\{1,\ldots,n\})$ represents the $i$-th sample, $Y_i\ (\in\mathbb{R})$ is the outcome variable, $T_i\ (\in\{0,1\})$ is the assignment variable, and $\bm{X}_i=(X_{1i},\ldots,X_{pi})'$ is the confounding variable vector for $Y_i$ and $T_i$. In addition, $\xi_i\ (\in\mathbb{R})$ is an unobserved random variable that depends on $\bm{X}_i$, and the conditional distribution given $\bm{X}_i$ is supposed to be normal. We assume that $T_i\indep Y_i \mid \bm{X}_i$, which is called the ignorable treatment assignment condition. In this model, the parameter of the causal effect is $\bm{\beta}^{(1)}-\bm{\beta}^{(0)}$, and the basic semiparametric method for consistent estimation is the inverse-probability-weighted estimation using the propensity score $e(\bm{X}_i)\equiv\P(T_i=1 \mid \bm{X}_i)$ (\citealt{Rub85}; \citealt{RobRZ94}). It has the advantage of not requiring an estimation of $\E(\xi_i\mid\bm{X}_i)$, which is difficult to model, and we rely upon it in this paper.

Now, supposing that the dimension $p$ of the confounding variable vector is high, let us try to estimate the causal effect by selecting variables to improve the estimation accuracy. Specifically, letting $\|\cdot\|_1$ be the $\ell_1$ norm, we consider
\begin{align*}
\sum_{i=1}^n\left[\left\{\frac{T_i}{e(\bm{X}_i)}-\frac{1-T_i}{1-e(\bm{X}_i)}\right\}Y_i-\bm{X}_i'\bm{\beta}\right]^2+\lambda\|\bm{\beta}\|_1
\end{align*}
and give its minimizer with respect to $\bm{\beta}$ as a Lasso-type estimator (\citealt{Tib96}). As mentioned at the beginning, this type of variable selection affects the subsequent inference of the causal effect, for example, in constructing a confidence interval for $\bm{\beta}^{(1)}-\bm{\beta}^{(0)}$. Therefore, we will attempt to combine causal inference with selective inference. Specifically, on the basis of the method of \cite{LeeSST16}, we can give a confidence interval with a confidence coefficient of $\alpha$ conditional on the event of variable selection for each component of $\bm{\beta}^{(1)}-\bm{\beta}^{(0)}$ with the guarantee of asymptotics. The same problem has been addressed by \cite{ZhaSE17}, who give a nonparametric estimation of $\E(\xi_i\mid\bm{X}_i)$ without using propensity scores. One of the reasons why propensity score analysis has become a hot topic is that it does not necessarily require this nonparametric estimation; we put much value on it in this paper. 

As a preparation, Section \ref{sec2} introduces propensity score analysis in causal inference models and selective inference with a focus on \cite{LeeSST16}. Section \ref{sec3} deals with selective inference in propensity score analysis and builds a foundation for selective inference by conditioning on the assignment variables before the outcome variables, an operation that is not usually considered in propensity score analysis. Then, asymptotically guaranteed confidence intervals are given by using higher-order asymptotic theory. Sections \ref{sec4} and \ref{sec5} describe numerical experiments and real data analyses comparing the proposed method with a naive method that ignores the influence of model selection. We find that there is a significant difference between the two methods and that the naive method lacks validity to a large extent. In Section \ref{sec6}, we extend the proposed method to handle especially developed versions of propensity score analysis and conclude in Section \ref{sec7}.

\section{Preparation}
\label{sec2}
\subsection{Causal inference model}
\label{sec2_1}
Let us assume that there are $H$ types of treatments, and let ${Y}^{(h)}\ (\in\mathbb{R})$ be the potential outcome variable when the $h$-th treatment is received. In addition, let $T^{(h)}$ be the assignment variable that is $1$ when $Y^{(h)}$ is observed and $0$ when it is not ($h\in\{1,\ldots,H\},\ \sum_{h=1}^HT^{(h)}=1$). Moreover, let $\bm{X}\ (\in\mathbb{R}^p)$ be the confounding variable vector, i.e., $p$ kinds of confounding variables are observed; we consider the model,
\begin{align}
{Y} \equiv \sum_{h=1}^HT^{(h)}{Y}^{(h)}
= \sum_{h=1}^HT^{(h)} \{ {\mu}^{(h)}(\bm{X})+{f}(\bm{X})+{\epsilon}^{(h)} \}.
\label{model1}
\end{align}
Here, $f:\mathbb{R}^p\to\mathbb{R}$ is a nonlinear function and ${\epsilon}^{(h)}$ is the error variable that follows a normal distribution ${\rm N}({0},{\sigma^2})$ independently of $\bm{X}$. Note that ${Y}$ is the observed outcome variable. Letting $(c^{(1)},\ldots,c^{(H)})'\ (\in\mathbb{R}^H)$ be a contrast satisfying $\sum_{h=1}^Hc^{(h)}=0$, we suppose that the object of interest is the conditional average treatment effect $\sum_{h=1}^H c^{(h)} {\mu}^{(h)}(\bm{x})$ for $\bm{X}=\bm{x}$. If $H=2$, and $h=1$ and $h=2$ represent the control and treatment groups, respectively, and if $\mu^{(1)}(\bm{x})$ and $\mu^{(2)}(\bm{x})$ do not depend on $\bm{x}$, then this model is the simplest Rubin's causal inference model, and the causal effect $\mu^{(2)}-\mu^{(1)}$ defined by $(c^{(1)},c^{(2)})'=(-1,1)'$ becomes the simple one estimated in \cite{RosR83} or \cite{Rub85}.

Suppose there are $n$ samples that follow this model and the variables in the $i$-th sample are identified by the subscript ${}_i$. Letting $\tilde{\bm{Y}}=({Y}_{1},\ldots,{Y}_{n})'$, $\tilde{\bm{T}}^{(h)}={\rm diag}(T_{1}^{(h)},\ldots,T_{n}^{(h)})$, $\tilde{\bm{Y}}^{(h)}=({Y}_{1}^{(h)},\ldots,{Y}_{n}^{(h)})'$, $\tilde{\bm{X}}=(\bm{X}_{1},\ldots,\bm{X}_{n})'$, $\tilde{\bm{\mu}}^{(h)}(\tilde{\bm{X}})=({\mu}^{(h)}(\bm{X}_1),\ldots,{\mu}^{(h)}(\bm{X}_n))'$, $\tilde{\bm{f}}(\tilde{\bm{X}})=({f}(\bm{X}_1),\allowbreak\ldots,{f}(\bm{X}_n))'$ and $\tilde{\bm{\epsilon}}^{(h)}=({\epsilon}_{1}^{(h)},\ldots,{\epsilon}_{n}^{(h)})'$, we express the model as
\begin{align}
\tilde{\bm{Y}} = \sum_{h=1}^{H} \tilde{\bm{T}}^{(h)} \tilde{\bm{Y}}^{(h)} = \sum_{h=1}^{H} \tilde{\bm{T}}^{(h)} \{ \tilde{\bm{\mu}}^{(h)}(\tilde{\bm{X}}) + \tilde{\bm{f}}(\tilde{\bm{X}}) + \tilde{\bm{\epsilon}}^{(h)} \}.
\label{model2}
\end{align}
We will assume the usual conditions for this setting. First, we assume the weakly ignorable treatment assignment condition (\citealt{Imb00}),
\begin{align}
{Y}_i^{(h)} \indep T_i^{(h)} \mid \bm{X}_i \qquad (i\in\{1,\ldots,n\},\ h\in\{1,\ldots,H\}).
\label{assump1}
\end{align}
Note that ${Y}_i^{(h)}$ in this condition can be changed to ${\epsilon}_i^{(h)}$. Second, we assume the positivity condition $0<\P(T_i^{(h)}=1\mid\bm{X}_i)<1$ ($i\in\{1,\ldots,n\}$). Third, we assume that the samples are independent, that is,
\begin{align*}
(T_{i}^{(h)},\bm{X}_{i},{\epsilon}_{i}^{(h)}) \indep (T_{l}^{(h)},\bm{X}_{l},{\epsilon}_{l}^{(h)}) \qquad (i\neq l, \ i,l\in\{1,\ldots,n\}, \ h\in\{1,\ldots,H\}).
\end{align*}
As a natural consequence, this means that ${Y}_{i}\indep {Y}_{l}$ ($i\neq l, \ i,l\in\{1,\ldots,n\}$).

In this paper, we will suppose that the dimension $p$ of the confounding variable vector is high and try to estimate the conditional average treatment effect with higher accuracy by simultaneously conducting variable selection and estimation. Specifically, we will use a Lasso-type method (\citealt{Tib96}) in Section \ref{sec3}. Note that a model selection method for causal inference was recently proposed by \cite{BabKN17}, but it deals with the selection of marginal structures and is not used here.


In \eqref{model2}, the $H-1$ potential outcome variables ${Y}_i^{(h)}$ with $T_i^{(h)}=0$ are considered to be missing. Since $\E({Y}_i^{(h)})\neq \E({Y}_i^{(h)}\mid T_i^{(h)}=1)$ in general, simply estimating the causal effect $\sum_{h=1}^Hc^{(h)}{\mu}^{(h)}(\bm{x})$ by using the least-squares method based on minimization of $\|\sum_{h=1}^H c^{(h)} \tilde{\bm{T}}^{(h)}\tilde{\bm{Y}}-\sum_{h=1}^H c^{(h)} \tilde{\bm{\mu}}^{(h)}(\tilde{\bm{X}})\|_2^2$ would result in a large bias. Here, $\|\cdot\|_2$ is the $\ell_2$ norm. As described in Section \ref{sec1}, this paper deals with inverse-probability-weighted estimation using the propensity score $e^{(h)}(\bm{X}_i)\equiv\P(T_i^{(h)}=1\mid\bm{X}_i)$ to avoid the bias.

In this estimation method, the missing values are pseudo-recovered by multiplying the observed values by the inverse of the propensity score as weights; then, the usual estimation is conducted. Specifically, letting $\tilde{\bm{W}}^{(h)}(\tilde{\bm{T}}^{(h)},\tilde{\bm{X}}) \equiv {\rm diag}\{T^{(h)}_1/e^{(h)}(\bm{X}_1),\ldots,T^{(h)}_n/e^{(h)}(\bm{X}_n)\}$ and $\tilde{\bm{T}}=(\tilde{\bm{T}}^{(1)},\ldots,\tilde{\bm{T}}^{(H)})$ and defining $\tilde{\bm{W}}(\tilde{\bm{T}},\tilde{\bm{X}}) \equiv \sum_{h=1}^{H} c^{(h)}\tilde{\bm{W}}^{(h)}(\tilde{\bm{T}}^{(h)},\tilde{\bm{X}})$ as the weight matrix, the inverse-probability-weighted estimator can be found by minimizing the squared loss function,
\begin{align}
\bigg\| \tilde{\bm{W}}(\tilde{\bm{T}},\tilde{\bm{X}}) \tilde{\bm{Y}} - \sum_{h=1}^{H} c^{(h)} \tilde{\bm{\mu}}^{(h)}(\tilde{\bm{X}}) \bigg\|_2^2.
\label{IPWloss}
\end{align}
When the propensity scores are unknown, we assume some parametric function for $e^{(h)}(\bm{X}_i)$, maximize the likelihood function $\prod_{i=1}^n \prod_{h=1}^H e^{(h)}(\bm{X}_i)^{T_i^{(h)}}$ to obtain the estimator $\hat{e}^{(h)}(\bm{X}_i)$, and use it instead. The inverse-probability-weighted estimator is consistent under the weakly ignorable treatment assignment condition in \eqref{assump1}.

\subsection{Selective inference}\label{sec2_3}

In Section \ref{sec2_1}, we suggested the use of variable selection. After the variable selection, in order to measure the extent to which the selected variables have an impact on the causal effects, tests are performed or confidence intervals are constructed.
However, the problem is that the $p$-values used in this process are no longer reliable, because the selected variables are likely to be significant, or in other words, the model using the selected variables likely overfits the data.

To explain the inference after variable selection in detail, we will omit the superscript ${}^{(h)}$ as $H=1$ in \eqref{model2}; i.e., $\tilde{\bm{T}}$ is an $n\times n$ identity matrix $\bm{I}_{n}$. Let $\tilde{\bm{X}}$ be a non-random matrix $\tilde{\bm{x}}$, and let $\tilde{\bm{f}}(\tilde{\bm{x}})$ be an $n$-dimensional zero vector $\bm{0}_{n}$. Moreover, we will consider a linear function of $\tilde{\bm{x}}$ as a model for $\tilde{\bm{\mu}}(\tilde{\bm{x}})$. Then, for the subset $\bm{M}\subset\{1,\ldots,p\}$, we define an estimand as
\begin{align*}
\bm{\beta}^{\ddagger\bm{M}} \equiv \argmin_{\bm{b}^{\ddagger\bm{M}}} \E ( \| \tilde{\bm{Y}}-\tilde{\bm{x}}_{\bm{M}}\bm{b}^{\ddagger\bm{M}} \|_2^2 ) = ( \tilde{\bm{x}}_{\bm{M}}' \tilde{\bm{x}}_{\bm{M}} )^{-1} \tilde{\bm{x}}_{\bm{M}}' \tilde{\bm{\mu}}(\tilde{\bm{x}}),
\end{align*}
which minimizes the expected squared error for the linear sum of the confounding variables belonging to $\bm{M}$, where $\tilde{\bm{x}}_{\bm{M}}=(\tilde{x}_{ij})_{i\in\{1,\ldots,n\},j\in\bm{M}}$. If the models $\bm{M}$ and $\bm{M}^*$ are different, then $\beta_j^{\ddagger\bm{M}}$ and $\beta_j^{\ddagger\bm{M}^*}$ are generally different, i.e., the target of inference is different for each model selected. Consequently, we can say that inference after selection has some ambiguity.

In terms of confidence interval construction, the simultaneous inference mentioned in Section \ref{sec1} is a method of creating an interval in which all the regression coefficients regardless of which model is selected are included with probability $1-\alpha$ or greater. When $\hat{\bm{M}}$ is the selected model and the regression coefficients in that model are denoted as $\beta_j^{\ddagger\hat{\bm{M}}}\ (j\in\hat{\bm{M}})$, this method finds $C_j^{\ddagger\hat{\bm{M}}}\ (j\in\hat{\bm{M}})$ such that
\begin{align*}
\P ( \forall j\in\hat{\bm{M}},\ \beta_j^{\ddagger\hat{\bm{M}}}\in C_j^{\ddagger\hat{\bm{M}}} ) \ge 1-\alpha.
\end{align*}
Since the interval is guaranteed for all $j$, a problem arises in that the interval becomes considerably wide when the size $|\hat{\bm{M}}|$ is large.

On the other hand, selective inference, in terms of confidence interval construction, is a method of creating an interval in which the regression coefficients in the selected model are included with probability $1-\alpha$ or greater under the condition that the model was selected. The method is to find $C_j^{\ddagger\bm{M}}\ (j\in\bm{M})$ such that
\begin{align*}
\P ( \beta_j^{\ddagger\bm{M}}\in C_j^{\ddagger\bm{M}} \mid \hat{\bm{M}}=\bm{M} ) \ge 1-\alpha.
\end{align*}
Accordingly, the false coverage rate is controlled to be
\begin{align}
\E \bigg( \frac{|\{j\in\hat{\bm{M}}:\beta_j^{\ddagger\hat{\bm{M}}}\notin C_j^{\ddagger\hat{\bm{M}}}\}|}{|\hat{\bm{M}}|};\ |\hat{\bm{M}}|>0 \bigg) \le \alpha.
\label{fcr}
\end{align}

In the cornerstone papers, beautiful inferences were presented for marginal screening and Lasso model selection in the setting of normal distributions. Then, their validity was theoretically demonstrated in more generalized settings. For example, \cite{FitST14} showed the optimality of the method in terms of uniformly most powerful unbiasedness in the setting of exponential families of distributions, while \cite{TiaT17}, \cite{TiaT18}, and \cite{TibRTW18} showed that the distributions of the pivot statistics for getting $p$-values or confidence intervals converge uniformly to the actual distribution when a normal distribution is used, even when the noise has a non-normal distribution. Furthermore, \cite{LiuMT18} proposed to increase the power of the original method by reducing the over-conditioning. In this paper, we develop selective inference for causal inference on the basis of \cite{LeeSST16}.


We will explain the method of \cite{LeeSST16}. Let us denote the usual Lasso estimator by $\hat{\bm{\beta}}^{\ddagger}=(\hat{\beta}_1^{\ddagger},\ldots,\hat{\beta}_p^{\ddagger})'$, the collection of non-zero estimators by $\hat{\bm{\beta}}^{\ddagger\hat{\bm{M}}}$, and its sign by $\hat{\bm{s}}^{\ddagger\hat{\bm{M}}}={\rm sign}(\hat{\bm{\beta}}^{\ddagger\hat{\bm{M}}})$. In this case, from the Karuch-Kuhn-Tucker conditions, for any $\bm{s}\in\{-1,1\}^{|\bm{M}|}$, there exists an $n\times n$ matrix $\bm{A}(\bm{M},\bm{s})$ and an $n$-dimensional vector $\bm{b}(\bm{M},\bm{s})$ such that
\begin{align*}
\{ \hat{\bm{M}}^{\ddagger}=\bm{M},\ \hat{\bm{s}}^{\ddagger\hat{\bm{M}}}=\bm{s} \}
= \{ \bm{A}(\bm{M},\bm{s})\tilde{\bm{Y}} \le \bm{b}(\bm{M},\bm{s}) \}.
\end{align*}

Now, using an appropriate unit vector $\bm{e}_j\ (\in\mathbb{R}^{|\bm{M}|})$, we define $\tilde{\bm{\eta}}_j^{\ddagger} \equiv \tilde{\bm{x}}_{\bm{M}} (\tilde{\bm{x}}_{\bm{M}}' \tilde{\bm{x}}_{\bm{M}})^{-1} \bm{e}_j$. Since the target parameter can be written as $\beta_j^{\ddagger\bm{M}} = \tilde{\bm{\eta}}_j^{\ddagger}{}' \tilde{\bm{\mu}}(\tilde{\bm{x}})$, we use the statistic $\tilde{\bm{\eta}}_j^{\ddagger}{}'\tilde{\bm{Y}}$ to create its confidence interval. Then, by using the strong properties of a normal distribution, we obtain
\begin{align*}
F_{\beta_j^{\ddagger\bm{M}}, \sigma^2 \tilde{\bm{\eta}}_j^{\ddagger}{}' \tilde{\bm{\eta}}_j^{\ddagger}}^{[{\mathcal V}_{\bm{s},j}^-(\tilde{\bm{Z}}), {\mathcal V}_{\bm{s},j}^+(\tilde{\bm{Z}})]} ( \tilde{\bm{\eta}}_j^{\ddagger}{}'\tilde{\bm{Y}} ) \mid \{ \bm{A}(\bm{M},\bm{s})\tilde{\bm{Y}} \le \bm{b}(\bm{M},\bm{s}) \} \sim {\rm Unif}(0,1).
\end{align*}
Here, $F_{\mu,\sigma^2}^{[a,b]}(\cdot)$ denotes the cumulative distribution function of ${\rm TN}(\mu,\sigma^2,a,b)$, which is ${\rm N}(\mu,\sigma^2)$ truncated into the interval $[a,b]$, $\tilde{\bm{Z}}=\{\bm{I}_n-\tilde{\bm{\eta}}_j^{\ddagger} (\tilde{\bm{\eta}}_j^{\ddagger}{}' \tilde{\bm{\eta}}_j^{\ddagger})^{-1} \tilde{\bm{\eta}}_j^{\ddagger}{}'\} \tilde{\bm{Y}}$, 
$$
{\mathcal V}_{\bm{s},j}^-(\tilde{\bm{Z}})=\max_{k:(\bm{A}(\bm{M},\bm{s}) \tilde{\bm{\eta}}_j^{\ddagger} (\tilde{\bm{\eta}}_j^{\ddagger}{}' \tilde{\bm{\eta}}_j^{\ddagger})^{-1})_k<0}\{b(\bm{M},\bm{s})_k-(\bm{A}(\bm{M},\bm{s})\tilde{\bm{Z}})_k\}/(\bm{A}(\bm{M},\bm{s}) \tilde{\bm{\eta}}_j^{\ddagger} (\tilde{\bm{\eta}}_j^{\ddagger}{}' \tilde{\bm{\eta}}_j^{\ddagger})^{-1})_k,
$$
$$
{\mathcal V}_{\bm{s},j}^+(\tilde{\bm{Z}})=\min_{k:(\bm{A}(\bm{M},\bm{s}) \tilde{\bm{\eta}}_j^{\ddagger} (\tilde{\bm{\eta}}_j^{\ddagger}{}' \tilde{\bm{\eta}}_j^{\ddagger})^{-1})_k>0}\{b(\bm{M},\bm{s})_k-(\bm{A}\allowbreak(\bm{M},\bm{s})\tilde{\bm{Z}})_k\}/(\bm{A}(\bm{M},\bm{s}) \tilde{\bm{\eta}}_j^{\ddagger} (\tilde{\bm{\eta}}_j^{\ddagger}{}' \tilde{\bm{\eta}}_j^{\ddagger})^{-1})_k,
$$
and ${\rm Unif}(0,1)$ denotes a continuous uniform distribution in $[0,1]$. We have obtained a pivot statistic that is unique to a normal distribution, and from it we can construct a statistical test or a confidence interval for $\beta_j^{\ddagger\bm{M}}$, with the significance level or coverage probability controlled in a conditional sense.

\section{Main result}
\label{sec3}
Consider selective inference for the causal inference model in \eqref{model2}. This problem has been addressed by \cite{ZhaSE17}, who do not use propensity scores, but perform nonparametric estimation of $\tilde{\bm{f}}(\tilde{\bm{X}})$. One of the reasons why causal inference has become a hot topic is the rapid development of propensity score analysis, which does not necessarily require this nonparametric estimation, and so, in this section, we will avoid the nonparametric estimation as well. This avoidance would require an essential generalization of selective inference, unlike \cite{ZhaSE17}.

Now, for $\tilde{\bm{\mu}}^{(h)}(\tilde{\bm{X}})$ in \eqref{model2}, let us choose a linear function of $\tilde{\bm{X}}$ as a model. For the subset $\bm{M}\subset\{1,\ldots,p\}$, because the causal effect is $\sum_{h=1}^H c^{(h)}\mu^{(h)}(\bm{X})$, we define an estimand as
\begin{align*}
\bm{\beta}^{\bm{M}} \equiv \argmin_{\bm{b}^{\bm{M}}} \E \bigg( \bigg\| \sum_{h=1}^H c^{(h)} \tilde{\bm{Y}}^{(h)}-\tilde{\bm{X}}_{\bm{M}}\bm{b}^{\bm{M}} \bigg\|_2^2 \ \bigg|\ \tilde{\bm{X}} \bigg) = ( \tilde{\bm{X}}_{\bm{M}}' \tilde{\bm{X}}_{\bm{M}} )^{-1} \tilde{\bm{X}}_{\bm{M}}' \sum_{h=1}^H c^{(h)} \tilde{\bm{\mu}}^{(h)}(\tilde{\bm{X}}),
\end{align*}
which minimizes the expected squared error for the linear sum of the confounding variables belonging to $\bm{M}$, where $\tilde{\bm{X}}_{\bm{M}}=(\tilde{X}_{ij})_{i\in\{1,\ldots,n\}, j\in\bm{M}}$. Note that, in this model, the causal effect is captured by $\bm{x}_{\bm{M}}'\bm{\beta}^{\bm{M}}$. Then, denoting the selected model as $\hat{\bm{M}}$, we try to find $C_j^{\bm{M}}\ (j\in\bm{M})$, which gives the conditional coverage,
\begin{align}
\P ( \beta_j^{\bm{M}} \in C_j^{\bm{M}} \mid \hat{\bm{M}}=\bm{M} ) \ge 1-\alpha
\label{ccover}
\end{align}
for the regression coefficient $\beta_j^{\bm{M}}\ (j\in\bm{M})$.

This allows us to construct a statistical test or a confidence interval for $\beta_j^{\bm{M}}$. Using an appropriate unit vector $\bm{e}_j\ (\in\mathbb{R}^{|\bm{M}|})$, we have $\beta_j^{\bm{M}}=\bm{e}_j' (\tilde{\bm{X}}_{\bm{M}}' \tilde{\bm{X}}_{\bm{M}})^{- 1} \tilde{\bm{X}}_{\bm{M}}' \sum_{h=1}^H c^{(h)}\tilde{\bm{\mu}}^{(h)}(\tilde{\bm{X}})$. Then, defining $\tilde{\bm{\eta}}_j\equiv \tilde{\bm {X}}_{\bm{M}} (\tilde{\bm{X}}_{\bm{M}}' \tilde{\bm{X}}_{\bm{M}})^{-1} \bm{e}_j$, later we consider the distribution of $\tilde{\bm{\eta}}_j' \tilde{\bm{W}}(\tilde{\bm{T}},\tilde{\bm{X}}) \tilde{\bm{Y}}$ conditional on $\hat{\bm{M}}=\bm{M}$.
Since it is not possible to perfectly capture the true model by $\hat{\bm{M}}$ in finite data, and the true structure may not be included in the model, $\tilde{\bm{X}}_{\bm{M}}\bm{\beta}^{\bm{M}}$ is not itself the conditional average treatment effect $\sum_{h=1}^H c^{(h)}\tilde{\bm{\mu}}^{(h)}(\tilde{\bm{X}})$.  However, it is a problem with selective inference itself, and it is not addressed in this paper.

As a model selection method, by letting $\bm{\beta}=\sum_{h=1}^H c^{(h)} \bm{\beta}^{(h)}$ after using $\tilde{\bm{\mu}}^{(h)}(\tilde{\bm{X}})=\tilde{\bm{X}}\bm{\beta}^{(h)}$ in \eqref{IPWloss}, we propose a Lasso-type one that estimates $\bm{\beta}$ by
\begin{align}
\hat{\bm{\beta}} = \argmin_{\bm{\beta}} \bigg\{ \frac{1}{2} \| \tilde{\bm{W}} ( \tilde{\bm{T}},\tilde{\bm{X}} ) \tilde{\bm{Y}} - \tilde{\bm{X}} \bm{\beta} \|_2^2 + \lambda \| \bm{\beta} \|_1 \bigg\}.
\label{IPWlasso}
\end{align}
Note that $\tilde{\bm{Y}}$ has a nonlinear component $\tilde{\bm{f}}(\tilde{\bm{X}})$, however, this nonlinear component is canceled out in $\tilde{\bm{W}}(\tilde{\bm{T}},\tilde{\bm{X}})\tilde{\bm{Y}}$ and there is no problem in using Lasso for linear models.
Since Lasso gives a sparse solution, the selected model can be expressed as
\begin{align*}
\hat{\bm{M}} = \{j:\ \hat{\beta}_j \neq 0\}.
\end{align*}
Here, we write $\hat{\bm{\beta}}^{\hat{\bm{M}}} = (\hat{\beta}_j)_{j\in\hat{\bm{M}}}$ for the collection of non-zero estimators and $\hat{\bm{s}}^{\hat{\bm{M}}} = {\rm sign}(\hat{\bm{\beta}}^{\hat{\bm{M}}})$ for their sign.

The fact that $\hat{\bm{\beta}}$ is the solution of \eqref{IPWlasso} is equivalent to the existence of $\hat{\bm{s}}=(\hat{s}_1,\ldots,\hat{s}_p)'$ satisfying the Karush-Kuhn-Tucker conditions, 
\begin{align*}
& \hat{\beta}_j \neq 0 \quad \Rightarrow \quad \hat{s}_j = {\rm sign} ( \hat{\beta}_j ),
\\
& \hat{\beta}_j = 0 \quad \Rightarrow \quad \hat{s}_j \in [-1,1]
\end{align*}
and
\begin{align*}
\tilde{\bm{X}}' \{ \tilde{\bm{X}} \hat{\bm{\beta}}-\tilde{\bm{W}} ( \tilde{\bm{T}},\tilde{\bm{X}} ) \tilde{\bm{Y}} \}+\lambda \hat{\bm{s}}=\bm{0}.
\end{align*}
For the model $\bm{M}$, if the candidate for the sign of its regression coefficients is $\bm{s}\ (\in\{-1,1\}^{|\bm{M}|})$, then we can use this equivalence conditions to show
\begin{align}
\{ \hat{\bm{M}}=\bm{M},\ \hat{\bm{s}}^{\hat{\bm{M}}}=\bm{s} \} = \{ \bm{A}(\bm{M},\bm{s}) \tilde{\bm{W}} ( \tilde{\bm{T}},\tilde{\bm{X}} ) \tilde{\bm{Y}} \le \bm{b}(\bm{M},\bm{s}) \}.
\label{condi1}
\end{align}
Here, we use the definitions,
\begin{align*}
\bm{A}(\bm{M},\bm{s}) = \frac{1}{\lambda} \left( \begin{array}{c} \tilde{\bm{X}}_{\bm{M}^{\rm c}}' \{ \bm{I}_n-\tilde{\bm{X}}_{\bm{M}} ( \tilde{\bm{X}}_{\bm{M}}' \tilde{\bm{X}}_{\bm{M}} )^{-1} \tilde{\bm{X}}_{\bm{M}}' \} \\ -\tilde{\bm{X}}_{\bm{M}^{\rm c}}' \{ \bm{I}_n-\tilde{\bm{X}}_{\bm{M}} ( \tilde{\bm{X}}_{\bm{M}}' \tilde{\bm{X}}_{\bm{M}} )^{-1} \tilde{\bm{X}}_{\bm{M}}' \} \\ -{\rm diag}(\bm{s}) ( \tilde{\bm{X}}_{\bm{M}}' \tilde{\bm{X}}_{\bm{M}} )^{-1} \tilde{\bm{X}}_{\bm{M}}' \end{array} \right)
\end{align*}
and
\begin{align*}
\bm{b}(\bm{M},\bm{s}) = \left( \begin{array}{c} \bm{1}_{|\bm{M}^{\rm c}|}-\tilde{\bm{X}}_{\bm{M}^{\rm c}}' \tilde{\bm{X}}_{\bm{M}} ( \tilde{\bm{X}}_{\bm{M}}' \tilde{\bm{X}}_{\bm{M}} )^{-1} \bm{s} \\ \bm{1}_{|\bm{M}^{\rm c}|}+\tilde{\bm{X}}_{\bm{M}^{\rm c}}' \tilde{\bm{X}}_{\bm{M}} ( \tilde{\bm{X}}_{\bm{M}}' \tilde{\bm{X}}_{\bm{M}} )^{-1} \bm{s} \\ -{\rm diag}(\bm{s}) ( \tilde{\bm{X}}_{\bm{M}}' \tilde{\bm{X}}_{\bm{M}} )^{-1} \bm{s} \end{array} \right),
\end{align*}
where $\bm{M}^{\rm c}=\{1,\ldots,p\}\setminus\bm{M}$, $\tilde{\bm{\bm{X}}}_{\bm{M}^{\rm c}}=(\tilde{X}_{ij})_{i\in\{1,\ldots,n\},j\in\bm{M}^{\rm c}}$ and $\bm{1}_{|\bm{M}^{\rm c}|}$ is a $|\bm{M}^{\rm c}|$-dimensional one-vector. Its derivation method is the same as the one used by \cite{LeeSST16}. Hence, we also have
\begin{align}
\{ \hat{\bm{M}}=\bm{M} \} = \bigcup_{\bm{s}\in\{-1,1\}^{\bm{M}}} \{ \bm{A}(\bm{M},\bm{s}) \tilde{\bm{W}} ( \tilde{\bm{T}},\tilde{\bm{X}} ) \tilde{\bm{Y}} \le \bm{b}(\bm{M},\bm{s}) \}.
\label{condunion}
\end{align}

Although we mentioned that we would consider the conditional coverage in \eqref{ccover}, conditioning also on the sign $\hat{\bm{s}}^{\hat{\bm{M}}}$ leads to an easier problem; hence, we will consider
\begin{align*}
& \P ( \beta_j^{\bm{M}} \in C_j^{\bm{M}} \mid \hat{\bm{M}}=\bm{M},\ \hat{\bm{s}}^{\hat{\bm{M}}}=\bm{s} ) 
\\
& = \P \{ \beta_j^{\bm{M}} \in C_j^{\bm{M}} \mid \bm{A}(\bm{M},\bm{s}) \tilde{\bm{W}} ( \tilde{\bm{T}},\tilde{\bm{X}} ) \tilde{\bm{Y}} \le \bm{b}(\bm{M},\bm{s}) \} \ge 1-\alpha
\end{align*}
as in \cite{LeeSST16}. This interval $C_j^{\bm{M}}$ also satisfies \eqref{ccover}, but if one wants tighter coverage, one can condition on \eqref{condunion}. To find $C_j^{\bm{M}}$, we consider the distribution of $\tilde{\bm{\eta}}_j' \tilde{\bm{W}} ( \tilde{\bm{T}},\tilde{\bm{X}} ) \tilde{\bm{Y}}$ conditional on \eqref{condi1}. Let $\tilde{\bm{D}}(\tilde{\bm{T}},\tilde{\bm{X}}) \equiv \tilde{\bm{W}}(\tilde{\bm{T}},\tilde{\bm{X}})^2 \tilde{\bm{\eta}}_j \{\tilde{\bm{\eta}}_j' \tilde{\bm{W}}(\tilde{\bm{T}},\tilde{\bm{X}})^2 \tilde{\bm{\eta}}_j\}^{-1}$. Using the same derivation method as in \cite{LeeSST16}, we can rewrite the conditioned polyhedral region by using $\tilde{\bm{\eta}}_j' \tilde{\bm{W}}(\tilde{\bm{T}},\tilde{\bm{X}}) \tilde{\bm{Y}}$ and
\begin{align*}
\tilde{\bm{Z}} \equiv \{ \bm{I}_n-\tilde{\bm{D}}(\tilde{\bm{T}},\tilde{\bm{X}}) \tilde{\bm{\eta}}_j' \} \tilde{\bm{W}} ( \tilde{\bm{T}},\tilde{\bm{X}} ) \tilde{\bm{Y}},
\end{align*}
from which we get
\begin{align*}
& \{ \bm{A}(\bm{M},\bm{s}) \tilde{\bm{W}} ( \tilde{\bm{T}},\tilde{\bm{X}} ) \tilde{\bm{Y}} \le \bm{b}(\bm{M},\bm{s}) \} 
\\
& = \{ {\mathcal V}_{\bm{s},j}^- ( \tilde{\bm{Z}},\tilde{\bm{T}},\tilde{\bm{X}} ) \le \tilde{\bm{\eta}}_j' \tilde{\bm{W}} ( \tilde{\bm{T}},\tilde{\bm{X}} ) \tilde{\bm{Y}} \le {\mathcal V}_{\bm{s},j}^+ ( \tilde{\bm{Z}},\tilde{\bm{T}},\tilde{\bm{X}} ),\ {\mathcal V}_{\bm{s},j}^0 ( \tilde{\bm{Z}},\tilde{\bm{T}},\tilde{\bm{X}} ) \ge 0 \},
\end{align*}
where 
\begin{align}
& {\mathcal V}_{\bm{s},j}^- ( \tilde{\bm{Z}},\tilde{\bm{T}},\tilde{\bm{X}} ) \equiv \max_{k:(\bm{A}(\bm{M},\bm{s}) \tilde{\bm{D}}(\tilde{\bm{T}},\tilde{\bm{X}}))_k<0} \frac{\bm{b}(\bm{M},\bm{s})_k-(\bm{A}(\bm{M},\bm{s}) \tilde{\bm{Z}})_k}{(\bm{A}(\bm{M},\bm{s}) \tilde{\bm{D}}(\tilde{\bm{T}},\tilde{\bm{X}}))_k},
\label{Vdef1}
\\
& {\mathcal V}_{\bm{s},j}^+ ( \tilde{\bm{Z}},\tilde{\bm{T}},\tilde{\bm{X}} ) \equiv \min_{k:(\bm{A}(\bm{M},\bm{s}) \tilde{\bm{D}}(\tilde{\bm{T}},\tilde{\bm{X}}))_k>0} \frac{\bm{b}(\bm{M},\bm{s})_k-(\bm{A}(\bm{M},\bm{s}) \tilde{\bm{Z}})_k}{(\bm{A}(\bm{M},\bm{s}) \tilde{\bm{D}}(\tilde{\bm{T}},\tilde{\bm{X}}))_k}
\label{Vdef2}
\end{align}
and
\begin{align*}
{\mathcal V}_{\bm{s},j}^0 ( \tilde{\bm{Z}},\tilde{\bm{T}},\tilde{\bm{X}} ) \equiv \min_{k:(\bm{A}(\bm{M},\bm{s}) \tilde{\bm{D}}(\tilde{\bm{T}},\tilde{\bm{X}}))_k=0} \{ \bm{b}(\bm{M},\bm{s})_k- ( \bm{A}(\bm{M},\bm{s}) \tilde{\bm{Z}} )_k \}.
\end{align*}

Given this setup, we want to construct a statistical test or a confidence interval for the target parameter $\beta_j^{\bm{M}}=\bm{e}_j' (\tilde{\bm{X}}_{\bm{M}}' \tilde{\bm{X}}_{\bm{M}})^{-1} \tilde{\bm{X}}_{\bm{M}}' \sum_{h=1}^H c^{(h)} \allowbreak \tilde{\bm{\mu}}^{(h)}(\tilde{\bm{X}})$. However, unlike in Section \ref{sec2_3}, $\tilde{\bm{Y}}$ is not normally distributed and $\tilde{\bm{f}}(\tilde{\bm{X}})$ exists. Also, $\tilde{\bm{X}}$ is random, although it is a trivial difference. Therefore, we consider using the realizations $\tilde{\bm{t}}$ and $\tilde{\bm{x}}$ to further condition on that $\tilde{\bm{T}}=\tilde{\bm{t}}$ and $\tilde{\bm{X}}=\tilde{\bm{x}}$. This is a difficult operation to understand because, $\tilde{\bm{T}}$ is not conditioned first in the usual propensity score analysis, but it is essentially necessary to extract the properties of a normal distribution. It is also difficult to intuitively understand how such conditioning guarantees valid inferences, but as we will see later, it is easy to show that it does. In general, under arbitrary conditioning, inference about the target parameter becomes impossible, but in this setting, under the weakly ignorable treatment assignment condition, inference based on asymptotics is possible. Also, as in the usual propensity score analysis, it is independent of $\tilde{\bm{f}}(\tilde{\bm{X}})$.

Under this new conditioning, $\tilde{\bm{\eta}}_j' \tilde{\bm{W}}(\tilde{\bm{T}},\tilde{\bm{X}}) \tilde{\bm{Y}}$ and $\tilde{\bm{Z}}$ become independent; thus, letting $\tilde{\bm{z}}$ be the realization of $\tilde{\bm{Z}}$, we have
\begin{align}
& [ \tilde{\bm{\eta}}_j' \tilde{\bm{W}} ( \tilde{\bm{T}},\tilde{\bm{X}} ) \tilde{\bm{Y}} \mid \bm{A}(\bm{M},\bm{s}) \tilde{\bm{W}} ( \tilde{\bm{t}},\tilde{\bm{x}} ) \tilde{\bm{Y}} \le \bm{b}(\bm{M},\bm{s}),\ \tilde{\bm{Z}}=\tilde{\bm{z}},\ \tilde{\bm{T}}=\tilde{\bm{t}},\ \tilde{\bm{X}}=\tilde{\bm{x}} ]
\nonumber\\
& \stackrel{\rm d}{=} [ \tilde{\bm{\eta}}_j' \tilde{\bm{W}}( \tilde{\bm{t}},\tilde{\bm{x}} ) \tilde{\bm{Y}} \mid {\mathcal V}_{\bm{s},j}^- ( \tilde{\bm{z}},\tilde{\bm{t}},\tilde{\bm{x}} ) \le \tilde{\bm{\eta}}_j' \tilde{\bm{W}} ( \tilde{\bm{t}},\tilde{\bm{x}} ) \tilde{\bm{Y}} \le {\mathcal V}_{\bm{s},j}^+ ( \tilde{\bm{z}},\tilde{\bm{t}},\tilde{\bm{x}} ),\ \tilde{\bm{T}}=\tilde{\bm{t}},\ \tilde{\bm{X}}=\tilde{\bm{x}} ].
\label{condi2}
\end{align}
In addition, because the ignorable treatment assignment condition is satisfied, $\tilde{\bm{\epsilon}}^{(h)}$ follows ${\rm N}(\bm{0}_n,\allowbreak\sigma^2\bm{I}_n)$ under this conditioning. Hence, by letting 
\begin{align}
\kappa_j^{\bm{M}} ( \tilde{\bm{t}},\tilde{\bm{x}} ) \equiv \bm{e}_j' ( \tilde{\bm{x}}_{\bm{M}}' \tilde{\bm{x}}_{\bm{M}} )^{-1} \tilde{\bm{x}}_{\bm{M}}' \sum_{h=1}^H c^{(h)} \tilde{\bm{W}}^{(h)} ( \tilde{\bm{t}}^{(h)},\tilde{\bm{x}} ) \{ \tilde{\bm{\mu}}^{(h)}( \tilde{\bm{x}} ) + \tilde{\bm{f}} ( \tilde{\bm{x}} ) \}
\label{gamma}
\end{align}
and
\begin{align}
\zeta_j^{\bm{M}} ( \tilde{\bm{t}},\tilde{\bm{x}} ) \equiv \sigma^2 \bm{e}_j' ( \tilde{\bm{x}}_{\bm{M}}' \tilde{\bm{x}}_{\bm{M}} )^{-1} \tilde{\bm{x}}_{\bm{M}}' \tilde{\bm{W}} ( \tilde{\bm{t}},\tilde{\bm{x}} )^2 \tilde{\bm{x}}_{\bm{M}} ( \tilde{\bm{x}}_{\bm{M}}' \tilde{\bm{x}}_{\bm{M}} )^{-1} \bm{e}_j,
\label{zeta}
\end{align}
it can be seen that the conditional distribution is a truncated normal distribution,
\begin{align*}
{\rm TN} ( \kappa_j^{\bm{M}} ( \tilde{\bm{t}},\tilde{\bm{x}} ),\zeta_j^{\bm{M}} ( \tilde{\bm{t}},\tilde{\bm{x}} ), {\mathcal V}_{\bm{s},j}^- ( \tilde{\bm{z}},\tilde{\bm{t}},\tilde{\bm{x}} ), {\mathcal V}_{\bm{s},j}^+ ( \tilde{\bm{z}},\tilde{\bm{t}},\tilde{\bm{x}} ) ).
\end{align*}
Thus, letting $F^{\tilde{\bm{Z}},\tilde{\bm{T}},\tilde{\bm{X}}} (\tilde{\bm{\eta}}_j' \tilde{\bm{W}}(\tilde{\bm{T}},\tilde{\bm{X}}) \tilde{\bm{Y}}) \equiv F_{\kappa_j^{\bm{M}}(\tilde{\bm{T}},\tilde{\bm{X}}),\zeta_j^{\bm{M}}(\tilde{\bm{T}},\tilde{\bm{X}})}^{[{\mathcal V}_{\bm{s},j}^-(\tilde{\bm{Z}},\tilde{\bm{T}},\tilde{\bm{X}}),{\mathcal V}_{\bm{s},j}^+(\tilde{\bm{Z}},\tilde{\bm{T}},\tilde{\bm{X}})]} \allowbreak (\tilde{\bm{\eta}}_j' \tilde{\bm{W}}(\tilde{\bm{T}},\tilde{\bm{X}}) \tilde{\bm{Y}})$, we obtain 
\begin{align*}
& [ F^{\tilde{\bm{Z}},\tilde{\bm{T}},\tilde{\bm{X}}} ( \tilde{\bm{\eta}}_j' \tilde{\bm{W}} ( \tilde{\bm{T}},\tilde{\bm{X}} ) \tilde{\bm{Y}} ) \mid \bm{A}(\bm{M},\bm{s}) \tilde{\bm{W}} ( \tilde{\bm{T}},\tilde{\bm{X}} ) \tilde{\bm{Y}} \le \bm{b}(\bm{M},\bm{s}),\ \tilde{\bm{Z}}=\tilde{\bm{z}},\ \tilde{\bm{T}}=\tilde{\bm{t}},\ \tilde{\bm{X}}=\tilde{\bm{x}} ] 
\\
& \stackrel{\rm d}{=} [ F^{\tilde{\bm{z}},\tilde{\bm{t}},\tilde{\bm{x}}} ( \tilde{\bm{\eta}}_j' \tilde{\bm{W}}( \tilde{\bm{t}},\tilde{\bm{x}} ) \tilde{\bm{Y}} ) \mid \bm{A}(\bm{M},\bm{s}) \tilde{\bm{W}} ( \tilde{\bm{t}},\tilde{\bm{x}} ) \tilde{\bm{Y}} \le \bm{b}(\bm{M},\bm{s}),\ \tilde{\bm{Z}}=\tilde{\bm{z}},\ \tilde{\bm{T}}=\tilde{\bm{t}},\ \tilde{\bm{X}}=\tilde{\bm{x}} ]
\\
& \sim {\rm Unif}(0,1).
\end{align*}
Denoting the conditional density function of $V$ by $p_{V}$, we have shown that 
$$
p_{F^{\tilde{\bm{Z}},\tilde{\bm{T}},\tilde{\bm{X}}} (\tilde{\bm{\eta}}_j' \tilde{\bm{W}}(\tilde{\bm{T}},\tilde{\bm{X}}) \tilde{\bm{Y}}) \mid \tilde{\bm{Z}},\tilde{\bm{T}},\tilde{\bm{X}}}(v \mid \tilde{\bm{z}},\tilde{\bm{t}},\tilde{\bm{x}})=1_{[0,1]}(v),
$$
from which it can be seen that
\begin{align*}
& p_{F^{\tilde{\bm{Z}},\tilde{\bm{T}},\tilde{\bm{X}}} (\tilde{\bm{\eta}}_j' \tilde{\bm{W}}(\tilde{\bm{T}},\tilde{\bm{X}}) \tilde{\bm{Y}})}(v) 
\\
& = \int \sum_{\tilde{\bm{t}}} \left\{ \int p_{F^{\tilde{\bm{Z}},\tilde{\bm{T}},\tilde{\bm{X}}} (\tilde{\bm{\eta}}_j' \tilde{\bm{W}}(\tilde{\bm{T}},\tilde{\bm{X}}) \tilde{\bm{Y}}) \mid \tilde{\bm{Z}},\tilde{\bm{T}},\tilde{\bm{X}}} \left( v \mid \tilde{\bm{z}},\tilde{\bm{t}},\tilde{\bm{x}} \right) p_{\tilde{\bm{Z}} \mid \tilde{\bm{T}},\tilde{\bm{X}}} \left( \tilde{\bm{z}} \mid \tilde{\bm{t}},\tilde{\bm{x}} \right) {\rm d}\tilde{\bm{z}} \right\} p_{\tilde{\bm{T}},\tilde{\bm{X}}} \left( \tilde{\bm{t}},\tilde{\bm{x}} \right) {\rm d}\tilde{\bm{x}}
\\
& = 1_{[0,1]}(v).
\end{align*}
This leads us to the following theorem.
\begin{theorem}
Let $F_{\mu,\sigma^2}^{[a,b]}(\cdot)$ be the cumulative distribution functions of ${\rm TN}(\mu,\sigma^2,a,b)$, which is ${\rm N}(\mu,\sigma^2)$ truncated into the interval $[a,b]$, and let ${\mathcal V}_{\bm{s},j}^-$, ${\mathcal V}_{\bm{s},j}^+$, $\kappa_j^{\bm{M}}$ and $\zeta_j^{\bm{M}}$ be the functions defined in \eqref{Vdef1}, \eqref{Vdef2}, \eqref{gamma} and \eqref{zeta}, respectively. Then, 
\begin{align*}
[ F_{\kappa_j^{\bm{M}}(\tilde{\bm{T}},\tilde{\bm{X}}),\zeta_j^{\bm{M}}(\tilde{\bm{T}},\tilde{\bm{X}})}^{[{\mathcal V}_{\bm{s},j}^-(\tilde{\bm{Z}},\tilde{\bm{T}},\tilde{\bm{X}}),{\mathcal V}_{\bm{s},j}^+(\tilde{\bm{Z}},\tilde{\bm{T}},\tilde{\bm{X}})]} ( \tilde{\bm{\eta}}_j' \tilde{\bm{W}} ( \tilde{\bm{T}},\tilde{\bm{X}} ) \tilde{\bm{Y}} ) \mid \bm{A}(\bm{M},\bm{s}) \tilde{\bm{W}} ( \tilde{\bm{T}},\tilde{\bm{X}} ) \tilde{\bm{Y}} \le \bm{b}(\bm{M},\bm{s}) ] \sim {\rm Unif}(0,1)
\end{align*}
holds.
\label{th1}
\end{theorem}
\noindent
The key point here is that, like the conditioning of $\tilde{\bm{Z}}$, the conditioning of $\tilde{\bm{T}}$ and $\tilde{\bm{X}}$ does not interfere with the inference.

This theorem cannot be used for the inference about $\beta_j^{\bm{M}}$ without further consideration, because unlike the selective inference in \cite{LeeSST16}, $\beta_j^{\bm{M}}$ does not appear explicitly in this pivot statistic.
Therefore, we consider higher-order asymptotic theory and extract $\beta_j^{\bm{M}}$ from $\kappa_j^{\bm{M}}(\tilde{\bm{T}},\tilde{\bm{X}})$, which appears as the mean parameter of the truncated normal distribution. 
Then, we construct an inference guaranteed asymptotically by evaluating the asymptotic behavior of $\kappa_j^{\bm{M}}(\tilde{\bm{T}},\tilde{\bm{X}})$.
Note that this operation is peculiar to the selective inference for propensity score analysis. 

In \eqref{gamma} of $(\tilde{\bm{t}},\tilde{\bm{x}})$ as $(\tilde{\bm{T}},\tilde{\bm{X}})$, we decompose the parts other than $\bm{e}_j' (\tilde{\bm{X}}_{\bm{M}}' \tilde{\bm{X}}_{\bm{M}})^{-1}$ and evaluate their expectations. Then, we obtain
\begin{align*}
\E \bigg[ \bm{X}_{\bm{M},i} \sum_{h=1}^H c^{(h)} \frac{T_i^{(h)}}{e^{(h)}(\bm{X}_i)} \{ \mu^{(h)}(\bm{X}_i) + f(\bm{X}_i) \} \ \bigg|\ \tilde{\bm{X}} \bigg] 
& = \bm{X}_{\bm{M},i} \sum_{h=1}^H c^{(h)} \{ \mu^{(h)}(\bm{X}_i) + f(\bm{X}_i) \} 
\\
& = \bm{X}_{\bm{M},i} \sum_{h=1}^H c^{(h)} \mu^{(h)}(\bm{X}_i),
\end{align*}
since the sum of contrasts is 0, where $\bm{X}_{\bm{M},i}=(X_{ij})_{j\in\bm{M}}$. 

Thus, we have
\begin{align*}
& \kappa_j^{\bm{M}} ( \tilde{\bm{T}},\tilde{\bm{X}} ) - \beta_j^{\bm{M}}
\\
& = \bm{e}_j' ( \tilde{\bm{X}}_{\bm{M}}' \tilde{\bm{X}}_{\bm{M}} )^{-1} \tilde{\bm{X}}_{\bm{M}}' \sum_{h=1}^H c^{(h)} \{ \tilde{\bm{W}}^{(h)} ( \tilde{\bm{T}}^{(h)},\tilde{\bm{X}} ) - \bm{I}_n \} \{ \tilde{\bm{\mu}}^{(h)}( \tilde{\bm{X}} ) + \tilde{\bm{f}} ( \tilde{\bm{X}} ) \} = \oP(1)
\end{align*}
from the law of large numbers. We prepare an estimator $Y_i^{(h)\dagger}$ for $\mu^{(h)}(\bm{x}_i)+f(\bm{x}_i)$, and let $\tilde{\bm{Y}}^{(h)\dagger}=(Y_1^{(h)\dagger},\ldots,Y_n^{(h)\dagger})'$ and $\tilde{\bm{Y}}^{\dagger}=(\tilde{\bm{Y}}^{(1)\dagger}{}',\ldots,\tilde{\bm{Y}}^{(H)\dagger}{}')'$. Accordingly, we define
\begin{align}
\tau_j^{\bm{M}} ( \tilde{\bm{Y}}^{\dagger},\tilde{\bm{T}},\tilde{\bm{X}} ) \equiv \bm{e}_j' ( \tilde{\bm{X}}_{\bm{M}}' \tilde{\bm{X}}_{\bm{M}} )^{-1} \tilde{\bm{X}}_{\bm{M}}' \sum_{h=1}^H c^{(h)} \{ \tilde{\bm{W}}^{(h)} ( \tilde{\bm{T}}^{(h)},\tilde{\bm{X}} ) - \bm{I}_n \} \tilde{\bm{Y}}^{(h)\dagger}. 
\label{tau}
\end{align}
This leads us to the following lemma.
\begin{lemma}
Let $F_{\mu,\sigma^2}^{[a,b]}(\cdot)$ be the cumulative distribution functions of ${\rm TN}(\mu,\sigma^2,a,b)$, which is ${\rm N}(\mu,\sigma^2)$ truncated into the interval $[a,b]$, and let ${\mathcal V}_{\bm{s},j}^-$, ${\mathcal V}_{\bm{s},j}^+$, $\tau_j^{\bm{M}}$ and $\zeta_j^{\bm{M}}$ be the functions defined in \eqref{Vdef1}, \eqref{Vdef2}, \eqref{tau} and \eqref{zeta}, respectively. Then, if $Y_i^{(h)\dagger}-\{\mu^{(h)}(\bm{X}_i)+f(\bm{X}_i)\}=\oP(1)$ under the condition, 
\begin{align*}
[ F_{\beta_j^{\bm{M}}+\tau_j^{\bm{M}}(\tilde{\bm{Y}}^{\dagger},\tilde{\bm{T}},\tilde{\bm{X}}),\zeta_j^{\bm{M}}(\tilde{\bm{T}},\tilde{\bm{X}})}^{[{\mathcal V}_{\bm{s},j}^-(\tilde{\bm{Z}},\tilde{\bm{T}},\tilde{\bm{X}}),{\mathcal V}_{\bm{s},j}^+(\tilde{\bm{Z}},\tilde{\bm{T}},\tilde{\bm{X}})]} ( \tilde{\bm{\eta}}_j' \tilde{\bm{W}} ( \tilde{\bm{T}},\tilde{\bm{X}} ) \tilde{\bm{Y}} ) \mid \bm{A}(\bm{M},\bm{s}) \tilde{\bm{W}} ( \tilde{\bm{T}},\tilde{\bm{X}} ) \tilde{\bm{Y}} \le \bm{b}(\bm{M},\bm{s}) ] \stackrel{\rm d}{\to} {\rm Unif}(0,1)
\end{align*}
holds as $n\to\infty$.
\label{th2}
\end{lemma}
\noindent
Note that although this lemma makes use of asymptotics unlike Theorem \ref{th1}, it is essentially different from those of \cite{TiaT17} and \cite{TibRTW18}. 
If we try to use the results of these papers simply because the error in the context of causal inference, $f(\bm{X})+\epsilon^{(h)}$, has a non-Gaussian distribution, the estimation will have a large bias due to the existence of confounding. 
On the other hand, in this paper, we assume a Gaussian distribution for $\epsilon^{(h)}$, and this lemma does not mean that our result is valid even if we make it non-Gaussian.

Similarly to \cite{LeeSST16}, if we condition on \eqref{condunion}, we only have to use the normal distribution ${\rm N}(\mu,\sigma^2)$ truncated into the union set $S$ of intervals. Writing its cumulative distribution function as $F^S_{\mu,\sigma^2}(\cdot)$, we get
\begin{align*}
& \bigg[ F_{\beta_j^{\bm{M}}+\tau_j^{\bm{M}}(\tilde{\bm{Y}}^{\dagger},\tilde{\bm{T}},\tilde{\bm{X}}),\zeta_j^{\bm{M}}(\tilde{\bm{T}},\tilde{\bm{X}})}^{\bigcup_{\bm{s}} [{\mathcal V}_{\bm{s},j}^-(\tilde{\bm{Z}},\tilde{\bm{T}},\tilde{\bm{X}}),{\mathcal V}_{\bm{s},j}^+(\tilde{\bm{Z}},\tilde{\bm{T}},\tilde{\bm{X}})]} ( \tilde{\bm{\eta}}_j' \tilde{\bm{W}} ( \tilde{\bm{T}},\tilde{\bm{X}} ) \tilde{\bm{Y}} ) \ \bigg|\ \bigcup_{\bm{s}} \{ \bm{A}(\bm{M},\bm{s}) \tilde{\bm{W}} ( \tilde{\bm{T}},\tilde{\bm{X}} ) \tilde{\bm{Y}} \le \bm{b}(\bm{M},\bm{s}) \} \bigg]
\\
& \stackrel{\rm d}{\to} {\rm Unif}(0,1).
\end{align*}
In addition, because $F_{\beta_j^{\bm{M}}+\tau_j^{\bm{M}}(\tilde{\bm{Y}}^{\dagger},\tilde{\bm{T}},\tilde{\bm{X}}),\zeta_j^{\bm{M}}(\tilde{\bm{T}},\tilde{\bm{X}})}^{\bigcup_{\bm{s}} [{\mathcal V}_{\bm{s},j}^-(\tilde{\bm{Z}},\tilde{\bm{T}},\tilde{\bm{X}}),{\mathcal V}_{\bm{s},j}^+(\tilde{\bm{Z}},\tilde{\bm{T}},\tilde{\bm{X}})]} (\tilde{\bm{\eta}}_j' \tilde{\bm{W}}(\tilde{\bm{T}},\tilde{\bm{X}}) \tilde{\bm{Y}})$ is a monotonically decreasing function with respect to $\beta_j^{\bm{M}}$, we can see that if we set $L$ and $U$ to satisfy
\begin{align*}
F_{L+\tau_j^{\bm{M}}(\tilde{\bm{Y}}^{\dagger},\tilde{\bm{T}},\tilde{\bm{X}}),\zeta_j^{\bm{M}}(\tilde{\bm{T}},\tilde{\bm{X}})}^{\bigcup_{\bm{s}} [{\mathcal V}_{\bm{s},j}^-(\tilde{\bm{Z}},\tilde{\bm{T}},\tilde{\bm{X}}),{\mathcal V}_{\bm{s},j}^+(\tilde{\bm{Z}},\tilde{\bm{T}},\tilde{\bm{X}})]} ( \tilde{\bm{\eta}}_j' \tilde{\bm{W}} ( \tilde{\bm{T}},\tilde{\bm{X}} ) \tilde{\bm{Y}} ) = 1-\frac{\alpha}{2}
\end{align*}
and
\begin{align*}
F_{U+\tau_j^{\bm{M}}(\tilde{\bm{Y}}^{\dagger},\tilde{\bm{T}},\tilde{\bm{X}}),\zeta_j^{\bm{M}}(\tilde{\bm{T}},\tilde{\bm{X}})}^{\bigcup_{\bm{s}} [{\mathcal V}_{\bm{s},j}^-(\tilde{\bm{Z}},\tilde{\bm{T}},\tilde{\bm{X}}),{\mathcal V}_{\bm{s},j}^+(\tilde{\bm{Z}},\tilde{\bm{T}},\tilde{\bm{X}})]} ( \tilde{\bm{\eta}}_j' \tilde{\bm{W}} ( \tilde{\bm{T}},\tilde{\bm{X}}) \tilde{\bm{Y}} ) = \frac{\alpha}{2},
\end{align*}
we get
\begin{align*}
\P ( \beta_j^{\bm{M}} \in [L,U] \mid \hat{\bm{M}}=\bm{M} ) \to 1-\alpha,
\end{align*}
which gives the asymptotically guaranteed conditional coverage.

Next, we develop a theory using a concrete and easily conceivable estimator for $\mu^{(h)}(\bm{X}_i)+f(\bm{X}_i)$. Suppose we have a sequence of real numbers $\{\delta_n\}$ that converges to 0, and let ${\mathcal N}_{i}^{(h)} \equiv \{ l\neq i: \| \bm{X}_l-\bm{X}_i \|_2 \le \delta_n,\ T_l^{(h)}=1\}$. We take $\delta_n$ such that ${\rm O}(1)\neq{\mathcal N}_{i}^{(h)}={\rm o}(n)$. For example, if the set of possible values of $\bm{X}_i$ is a closed set of bounded open sets in $\mathbb{R}^p$, then we only have to take $\delta_n$ such that ${\rm O}(n^{-1})\neq\delta_n={\rm o}(1)$. Letting $\hat{\bm{\beta}}^{(h)} \equiv \argmin_{\bm{\beta}} \|\tilde{\bm{T}}^{(h)}(\tilde{\bm{Y}}-\tilde{\bm{X}} \bm{\beta})\|_2^2$ and $\tilde{\bm{Y}}^{(h)*} \equiv ( \bm{X}_1'\hat{\bm{\beta}}^{(h)}+\sum_{l\in{\mathcal N}_{1}^{(h)}} (Y_l-\bm{X}_l'\hat{\bm{\beta}}^{(h)})/|{\mathcal N}_{1}^{(h)}|,\ldots,\bm{X}_n'\hat{\bm{\beta}}^{(h)}+\sum_{l\in{\mathcal N}_{n}^{(h)}} (Y_l-\bm{X}_l'\hat{\bm{\beta}}^{(h)})/|{\mathcal N}_{n}^{(h)}|)'$, if $f(\cdot)$ is Lipschitz continuous, then it is obvious that $\kappa_j^{\bm{M}}(\tilde{\bm{T}},\tilde{\bm{X}}) - \{\beta_j^{\bm{M}} + \tau_j^{\bm{M}}(\tilde{\bm{Y}}^*,\tilde{\bm{T}},\tilde{\bm{X}})\} = \oP(n^{-1/2})$, where $\tau_j^{\bm{M}}$ is the function defined in \eqref{tau} and $\tilde{\bm{Y}}^*=(\tilde{\bm{Y}}^{(1)*}{}',\ldots,\tilde{\bm{Y}}^{(H)*}{}')'$. Thus, we obtain 
\begin{align}
\E \{ \tilde{\bm{\eta}}_j' \tilde{\bm{W}} ( \tilde{\bm{T}},\tilde{\bm{X}} ) \tilde{\bm{Y}} - \tau_j^{\bm{M}} ( \tilde{\bm{Y}}^*,\tilde{\bm{T}},\tilde{\bm{X}} ) \mid \tilde{\bm{T}},\tilde{\bm{X}} \} = \beta_j^{\bm{M}} + \oP(n^{-1/2}).
\label{Eeval}
\end{align}
For the evaluation of the variance,
\begin{align*}
& \V \{ \tilde{\bm{\eta}}_j' \tilde{\bm{W}} ( \tilde{\bm{T}},\tilde{\bm{X}} ) \tilde{\bm{Y}}-\tau_j^{\bm{M}} ( \tilde{\bm{Y}}^*,\tilde{\bm{T}},\tilde{\bm{X}} ) \mid \tilde{\bm{T}},\tilde{\bm{X}} \} 
\\
& = \E ( [ \{ \tilde{\bm{\eta}}_j' \tilde{\bm{W}} ( \tilde{\bm{T}},\tilde{\bm{X}} ) \tilde{\bm{Y}} - \kappa_j^{\bm{M}} ( \tilde{\bm{T}},\tilde{\bm{X}} ) \} 
- \tau_j^{\bm{M}} ( \tilde{\bm{\epsilon}}^*,\tilde{\bm{T}},\tilde{\bm{X}} ) ]^2 \mid \tilde{\bm{T}},\tilde{\bm{X}} ),
\end{align*}
the conditional expectation of the square of the first term is $\zeta_j^{\bm{M}}(\tilde{\bm{T}},\tilde{\bm{X}})$, the conditional expectation of the square of the second term is
\begin{align}
\tilde{\bm{\eta}}_j' \sum_{h=1}^H \sum_{k=1}^H c^{(h)} c^{(k)} \E [ \{ \tilde{\bm{W}}^{(h)} ( \tilde{\bm{T}}^{(h)},\tilde{\bm{X}} ) - \bm{I}_n \} \tilde{\bm{\epsilon}}^{(h)*} \tilde{\bm{\epsilon}}^{(k)*}{}' \{ \tilde{\bm{W}}^{(k)} ( \tilde{\bm{T}}^{(k)},\tilde{\bm{X}} ) - \bm{I}_n \} \mid \tilde{\bm{T}},\tilde{\bm{X}} ] \tilde{\bm{\eta}}_j,
\label{CondE1}
\end{align}
and the conditional expectation of the product of the first and second terms is
\begin{align}
\tilde{\bm{\eta}}_j' \sum_{h=1}^H \sum_{k=1}^H c^{(h)} c^{(k)} \E [ \tilde{\bm{W}}^{(h)} ( \tilde{\bm{T}}^{(h)},\tilde{\bm{X}} ) \tilde{\bm{\epsilon}}^{(h)} \tilde{\bm{\epsilon}}^{(k)*}{}' \{ \tilde{\bm{W}}^{(k)} ( \tilde{\bm{T}}^{(k)},\tilde{\bm{X}} ) - \bm{I}_n \} \mid \tilde{\bm{T}},\tilde{\bm{X}} ] \tilde{\bm{\eta}}_j,
\label{CondE2}
\end{align}
where $\tilde{\bm{\epsilon}}^*=\tilde{\bm{Y}}^*-\E(\tilde{\bm{Y}}^* \mid \tilde{\bm{T}},\tilde{\bm{X}})$.

In \eqref{CondE1}, if $h\neq k$, then $\tilde{\bm{\epsilon}}^{(h)*}$ and $\tilde{\bm{\epsilon}}^{(k)*}$ are independent, and if $h=k$, then $\tilde{\bm{\epsilon }}^{(h)*}\tilde{\bm{\epsilon}}^{(k)*}{}'$ has at most $\sum_{i=1}^{n}|{\mathcal N}_{i}^{(h)}|^2$ non-zero components. Given the expectation with respect to $\tilde{\bm{T}}$ of the conditional expectation, the off-diagonal components of this matrix are 0 due to the weakly ignorable treatment assignment condition and the independence of $\{T_i^{(h)}\}$; moreover, the diagonal components are $(\sigma^2/|{\mathcal N}_{i}^{(h)}|)\{1-e^{(h)}(\bm{X}_i)\}/e^{(h)}(\bm{X}_i)$ from the assignment condition. Hence, we can see that \eqref{CondE1} becomes
\begin{align}
\sigma^2 \sum_{h=1}^H c^{(h)2} \tilde{\bm{\eta}}_j' \tilde{\bm{K}}^{(h)}(\tilde{\bm{X}}) \tilde{\bm{\eta}}_j + \oP(n^{-3/2}),
\label{CondE3}
\end{align}
where $\tilde{\bm{K}}^{(h)}(\tilde{\bm{X}})\equiv{\rm diag}[\{1/e^{(h)}(\bm{X}_1)-1\}/|{\mathcal N}_1^{(h)}|,\ldots,\{1/e^{(h)}(\bm{X}_n)-1\}/|{\mathcal N}_n^{(h)}|]$. In \eqref{CondE2}, if $h\neq k$, then $\tilde{\bm{\epsilon}}^{(h)}$ and $\tilde{\bm{\epsilon}}^{(k)*}$ are independent, and if $h=k$, then $\tilde{\bm{\epsilon }}^{(h)}\tilde{\bm{\epsilon}}^{(k)*}{}'$ has at most $\sum_{i=1}^{n}|{\mathcal N}_{i}^{(h)}|$ non-zero components. Given the expectation with respect to $\tilde{\bm{T}}$ of the conditional expectation, the off-diagonal components of this matrix are 0 due to the weakly ignorable treatment assignment condition and the independence of $\{T_i^{(h)}\}$. Moreover, the diagonal components are also 0 because of the assignment condition and the independence of $\epsilon_i^{(h)}$ and $\epsilon_i^{(h)*}$. Hence, we can see that \eqref{CondE2} becomes $\oP(n^{-3/2})$.

Thus, by defining 
\begin{align}
\rho_j^{\bm{M}} ( \tilde{\bm{T}},\tilde{\bm{X}} ) \equiv \zeta_j^{\bm{M}} ( \tilde{\bm{T}},\tilde{\bm{X}} ) + \sigma^2 \sum_{h=1}^H c^{(h)2} \tilde{\bm{\eta}}_j' \tilde{\bm{K}}^{(h)}(\tilde{\bm{X}}) \tilde{\bm{\eta}}_j, 
\label{rho}
\end{align}
it can be seen that 
\begin{align}
\V \{ \tilde{\bm{\eta}}_j' \tilde{\bm{W}} ( \tilde{\bm{T}},\tilde{\bm{X}} ) \tilde{\bm{Y}}-\tau_j^{\bm{M}} ( \tilde{\bm{Y}}^*,\tilde{\bm{T}},\tilde{\bm{X}} ) \mid \tilde{\bm{T}},\tilde{\bm{X}} \} = \rho_j^{\bm{M}} ( \tilde{\bm{T}},\tilde{\bm{X}} ) + \oP(n^{-3/2}).
\label{Veval}
\end{align}
From \eqref{Eeval} and \eqref{Veval}, we obtain the following theorem, which is an asymptotic version of \eqref{fcr}.
\begin{theorem}
Let $F_{\mu,\sigma^2}^{[a,b]}(\cdot)$ be the cumulative distribution functions of ${\rm TN}(\mu,\sigma^2,a,b)$, which is ${\rm N}(\mu,\sigma^2)$ truncated into the interval $[a,b]$, and let ${\mathcal V}_{\bm{s},j}^-$, ${\mathcal V}_{\bm{s},j}^+$, $\tau_j^{\bm{M}}$ and $\rho_j^{\bm{M}}$ be the functions defined in \eqref{Vdef1}, \eqref{Vdef2}, \eqref{tau} and \eqref{rho}, respectively. Then, if we define $L_j^{\hat{\bm{M}}}$ and $U_j^{\hat{\bm{M}}}$ such that
\begin{align*}
F_{L_j^{\hat{\bm{M}}}+\tau_j^{\hat{\bm{M}}}(\tilde{\bm{Y}}^*,\tilde{\bm{T}},\tilde{\bm{X}}),\rho_j^{\hat{\bm{M}}}(\tilde{\bm{T}},\tilde{\bm{X}})}^{\bigcup_{\bm{s}} [{\mathcal V}_{\bm{s},j}^-(\tilde{\bm{Z}},\tilde{\bm{T}},\tilde{\bm{X}}),{\mathcal V}_{\bm{s},j}^+(\tilde{\bm{Z}},\tilde{\bm{T}},\tilde{\bm{X}})]} ( \tilde{\bm{\eta}}_j' \tilde{\bm{W}} ( \tilde{\bm{T}},\tilde{\bm{X}} ) \tilde{\bm{Y}} ) = 1-\frac{\alpha}{2}
\end{align*}
and
\begin{align*}
F_{U_j^{\hat{\bm{M}}}+\tau_j^{\hat{\bm{M}}}(\tilde{\bm{Y}}^*,\tilde{\bm{T}},\tilde{\bm{X}}),\rho_j^{\hat{\bm{M}}}(\tilde{\bm{T}},\tilde{\bm{X}})}^{\bigcup_{\bm{s}} [{\mathcal V}_{\bm{s},j}^-(\tilde{\bm{Z}},\tilde{\bm{T}},\tilde{\bm{X}}),{\mathcal V}_{\bm{s},j}^+(\tilde{\bm{Z}},\tilde{\bm{T}},\tilde{\bm{X}})]} ( \tilde{\bm{\eta}}_j' \tilde{\bm{W}} ( \tilde{\bm{T}},\tilde{\bm{X}}) \tilde{\bm{Y}} ) = \frac{\alpha}{2},
\end{align*}
then
\begin{align*}
\E \bigg( \frac{|\{j\in\hat{\bm{M}}:\beta_j^{\hat{\bm{M}}}\notin [L_j^{\hat{\bm{M}}},U_j^{\hat{\bm{M}}}]\}|}{|\hat{\bm{M}}|};\ |\hat{\bm{M}}|>0 \bigg) \le \alpha
\end{align*}
holds as $n\to\infty$.
\label{th3}
\end{theorem}
\noindent
The key point here is that the pivot statistic does not depend on the nonparametric function $\tilde{\bm{f}}(\tilde{\bm{X}})$.

\begin{remark}
Regarding $\tilde{\bm{Y}}^*$, for example, when $H=2$ and $(c^{(1)},c^{(2)})=(-1,1)$, if we define $\tilde{\bm{Y}}^{(1)*} \equiv (\bm{X}_1'\hat{\bm{\beta}}^{(1)} + \{\sum_{l\in{\mathcal N}_{1}^{(1)}} (Y_l-\bm{X}_l'\hat{\bm{\beta}}^{(1)}) + \sum_{l\in{\mathcal N}_{1}^{(2)}} (Y_l-\bm{X}_{\hat{\bm{M}},l}'\hat{\bm{\beta}}^{\hat{\bm{M}}}-\bm{X}_l'\hat{\bm{\beta}}^{(1)})\}/(|{\mathcal N}_{1}^{(1)}|+|{\mathcal N}_{1}^{(2)}|),\ldots,\bm{X}_n'\hat{\bm{\beta}}^{(1)} + \{\sum_{l\in{\mathcal N}_{n}^{(1)}} (Y_l-\bm{X}_l'\hat{\bm{\beta}}^{(1)}) + \sum_{l\in{\mathcal N}_{n}^{(2)}} (Y_l-\bm{X}_{\hat{\bm{M}},l}'\hat{\bm{\beta}}^{\hat{\bm{M}}}-\bm{X}_l'\hat{\bm{\beta}}^{(1)})\}/(|{\mathcal N}_{n}^{(1)}|+|{\mathcal N}_{n}^{(2)}|))'$ and $\tilde{\bm{Y}}^{(2)*} \equiv (\bm{X}_1'\hat{\bm{\beta}}^{(2)} + \{\sum_{l\in{\mathcal N}_{1}^{(1)}} (Y_l+\bm{X}_{\hat{\bm{M}},l}'\hat{\bm{\beta}}^{\hat{\bm{M}}}-\bm{X}_l'\hat{\bm{\beta}}^{(2)}) + \sum_{l\in{\mathcal N}_{1}^{(2)}} (Y_l-\bm{X}_l'\hat{\bm{\beta}}^{(2)})\}/(|{\mathcal N}_{1}^{(1)}|+|{\mathcal N}_{1}^{(2)}|),\ldots,\bm{X}_n'\hat{\bm{\beta}}^{(2)} + \{\sum_{l\in{\mathcal N}_{n}^{(1)}} (Y_l+\bm{X}_{\hat{\bm{M}},l}'\hat{\bm{\beta}}^{\hat{\bm{M}}}-\bm{X}_l'\hat{\bm{\beta}}^{(2)}) + \sum_{l\in{\mathcal N}_{n}^{(2)}} (Y_l-\bm{X}_l'\hat{\bm{\beta}}^{(2)})\}/(|{\mathcal N}_{n}^{(1)}|+|{\mathcal N}_{n}^{(2)}|))'$, the accuracy tends to increase. Note that in the definition of $\tilde{\bm{K}}^{(h)}(\tilde{\bm{X}})$, $|{\mathcal N}_{i}^{(h)}|$ becomes $|{\mathcal N}_{i}^{(1)}|+|{\mathcal N}_{i}^{(2)}|$.
\end{remark}

\section{Numerical experiment}
\label{sec4}
We numerically verified the usefulness of the proposed method. Letting $H=2$, for the confounding variable $\bm{X}\ (\in\mathbb{R}^p)$, the outcome variable $Y\ (\in\mathbb{R})$ was observed through the model, 
\begin{align*}
Y = T \mu(\bm{X}) + f(\bm{X}) + \epsilon.
\end{align*}
That is, the treatment group was represented by $T=1$ and the control group by $T=0$. The error variable $\epsilon$ was assumed to follow a normal distribution ${\rm N}(0,\sigma^2)$ independently of $\bm{X}$, and the variance was assumed to be known. This model is a special case of \eqref{model1}, where the causal effect is $\mu(\bm{x})$.

The setup of the numerical experiment is as follows. In each run, we choose one of the following
\begin{description}
\item[\rm \quad (F1) \quad] $f(\bm{x}) = 0$
\item[\rm \quad (F2) \quad] $f(\bm{x})=3x_1 + x_2 + x_3 + x_4 + x_5 -3.5$
\item[\rm \quad (F3) \quad] $f(\bm{x})=x_1 + 0.5x_2 + 0.5x_3 + \pi\sin(\pi x_4)/32 + \pi\sin(\pi x_5)/32 - 1.125$
\end{description}
as an infinite-dimensional nuisance parameter $f(\cdot)$. In addition, we chose one of the following
\begin{description}
\item[\rm \quad (E1) \quad] $e(\bm{x})=0.5$
\item[\rm \quad (E2) \quad] $e(\bm{x}) = 1/\{1+\exp(- x_1 - 0.5 x_2 - 0.5 x_3 - 0.5 x_4 - 0.5 x_5 + 1.5)\}$
\end{description}
as the propensity score $e(\cdot)$ and one of the following
\begin{description}
\item[\rm \quad (M1) \quad] $\mu(\bm{x})=0$
\item[\rm \quad (M2) \quad] $\mu(\bm{x}) = 3x_1 + x_2 + x_3 + x_4 + x_5 -3.5$
\end{description}
as the causal effect $\mu(\cdot)$. Note that (E1) indicates a completely random assignment. In the case of (M1), all variables are inactive for $\mu(\bm{x})$, while in the case of (M2), the variables $\{x_1,x_2,x_3,x_4,x_5\}$ are active for $\mu(\bm{x})$. Each component of the confounding variable vector independently follows a continuous uniform distribution ${\rm U}(0, 1)$ or Bernoulli distribution ${\rm Ber}(1/2)$, and the data are generated from the above model with a sample size of $n=1000$. Here, we set $p$ to $5$ or $ 25$ and the error variance $\sigma^2$ to $0.25^2$.

We compared the proposed method (SI) with a method that ignores the influence of model selection (Naive). Model selection in Naive is conducted by using Lasso and then confidence intervals are constructed using a normal distribution instead of a truncated normal distribution. Because Naive ignores the fact that the selected variables are likely to be significant, it is supposed that it will have a large number of false positives and the validity of its inference will be impaired.

We evaluated the size $|\hat{M}|$ of the model selected by Lasso and the false coverage rate (FCR) by conducting 1,000 Monte Carlo simulations. We also evaluated the probability that $\beta_1,\ldots,\beta_5$ is significant, i.e., the probability that $\beta_1=0,\ldots,\beta_5=0$ is not covered in the confidence interval when $p=5$, and numbers of true positive (TP) and false positive (FP) when $p=25$. In all cases, the significance level was set to 0.05; i.e., the confidence coefficient was set to 0.95. Therefore, the closer the FCR is to 0.05, the more valid the method is. Note that with $\hat{M}$ as the selected model and $C_j^{\hat{\bm{M}}}$ as the confidence interval for $\beta_j^{\hat{\bm{M}}}$, TP is the number of intervals that did not cover zero when $\beta_j^{\hat{\bm{M}}}$ is non-zero, $|\{j\in\hat{\bm{M}}: \beta_j^{\hat{\bm{M}}}\neq 0,\ 0\notin C_j^{\hat{\bm{M}}}\}|$, and FP is the number intervals that did not cover zero when $\beta_j^{\hat{\bm{M}}}$ is zero, $|\{j\in\hat{\bm{M}}: \beta_j^{\hat{\bm{M}}}=0,\ 0\notin C_j^{\hat{\bm{M}}}\}|$. 

\subsection{Case of $p=5$}
\label{sec4_1}
First, let us consider (M1), where all variables are inactive. The confounding variables followed a Bernoulli distribution, and the sequence $\delta_n$ used to define the neighborhood ${\cal N}_i^{(h)}$ was set to $0$. The tuning parameter of Lasso was $\lambda=\sigma n^{-1/2}(\log p)^{1/2}$ when the infinite-dimensional nuisance parameter was (F1), and $\lambda=2\sigma n^{-1/2}(\log p)^{1/2}$ when it was (F2) or (F3). Table \ref{t:1} shows $|\hat{M}|$, the non-coverage probability, and the FCR. Since $\beta_j=0$ is correct for all variables, the closer the non-coverage probability is to 0.05, the better. Although the non-coverage probability of SI seems to be somewhat large when the propensity score and the nuisance parameter are (E2) and (F2), it is approximately 0.05 in most other cases. On the other hand, if we look at Naive, we can see that when the nuisance parameter is (F2) or (F3), 0 is not covered for all the selected variables, and the validity of the inference is impaired. Note that when the nuisance parameter is (F1), the non-coverage probability is greater than 0.05, although this value is not as extreme as when the nuisance parameter is (F2) or (F3). The FCR shows a similar trend, where SI keeps the FCR at almost 0.05, but Naive results in a very large value. 

\begin{table}[t!]
\caption{Case in which $p=5$, $X_{ij}\sim{\rm Ber}(1/2)$ and (M1) is correct. The figures in parentheses are standard deviations. In all settings, SI controls the FCR appropriately, but the inference of Naive is invalid. }
\centering
{\small
\begin{tabular}{cccccccccc}
\toprule
$e(\bm{x})$ & $f(\bm{x})$ & $|\hat{M}|$ & & $\beta_1$ & $\beta_2$ & $\beta_3$ & $\beta_4$ & $\beta_5$ & FCR \\ \midrule
& & & \multirow{2}{*}{SI} & 0.039 & 0.031 & 0.026 & 0.045 & 0.047 & 0.035 \\
& \multirow{2}{*}{(F1)} & \multirow{2}{*}{2.593 {\footnotesize (1.164)}} & & {\footnotesize (0.194)} & {\footnotesize (0.175)} & {\footnotesize (0.159)} & {\footnotesize (0.208)} & {\footnotesize (0.212)} & {\footnotesize (0.124)} \\
& & & \multirow{2}{*}{Naive} & 0.080 & 0.085 & 0.070 & 0.089 & 0.088 & 0.079 \\
& & & & {\footnotesize (0.272)} & {\footnotesize (0.279)} & {\footnotesize (0.255)} & {\footnotesize (0.285)} & {\footnotesize (0.284)} & {\footnotesize (0.178)} \\ \cmidrule(r){2-10}
& & & \multirow{2}{*}{SI} & 0.049 & 0.055 & 0.038 & 0.050 & 0.059 & 0.051 \\
\multirow{2}{*}{(E1)} & \multirow{2}{*}{(F2)} & \multirow{2}{*}{3.641 {\footnotesize (1.012)}} & & {\footnotesize (0.215)} & {\footnotesize (0.227)} & {\footnotesize (0.190)} & {\footnotesize (0.218)} & {\footnotesize (0.235)} & {\footnotesize (0.123)} \\
& & & \multirow{2}{*}{Naive} & 1.000 & 1.000 & 1.000 & 1.000 & 1.000 & 0.999 \\
& & & & {\footnotesize (0.000)} & {\footnotesize (0.000)} & {\footnotesize (0.000)} & {\footnotesize (0.000)} & {\footnotesize (0.000)} & {\footnotesize (0.032)} \\ \cmidrule(r){2-10}
& & & \multirow{2}{*}{SI} & 0.040 & 0.051 & 0.042 & 0.043 & 0.048 & 0.039 \\
& \multirow{2}{*}{(F3)} & \multirow{2}{*}{1.675 {\footnotesize (1.141)}} & & {\footnotesize (0.196)} & {\footnotesize (0.219)} & {\footnotesize (0.201)} & {\footnotesize (0.204)} & {\footnotesize (0.213)} & {\footnotesize (0.156)} \\
& & & \multirow{2}{*}{Naive} & 1.000 & 1.000 & 1.000 & 1.000 & 1.000 & 0.842 \\
& & & & {\footnotesize (0.000)} & {\footnotesize (0.000)} & {\footnotesize (0.000)} & {\footnotesize (0.000)} & {\footnotesize (0.000)} & {\footnotesize (0.365)} \\
\midrule
& & & \multirow{2}{*}{SI} & 0.043 & 0.049 & 0.040 & 0.050 & 0.044 & 0.042 \\
& \multirow{2}{*}{(F1)} & \multirow{2}{*}{2.734 {\footnotesize (1.154)}} & & {\footnotesize (0.202)} & {\footnotesize (0.216)} & {\footnotesize (0.196)} & {\footnotesize (0.217)} & {\footnotesize (0.204)} & {\footnotesize (0.134)} \\
& & & \multirow{2}{*}{Naive} & 0.081 & 0.076 & 0.080 & 0.064 & 0.070 & 0.068 \\
& & & & {\footnotesize (0.274)} & {\footnotesize (0.265)} & {\footnotesize (0.272)} & {\footnotesize (0.246)} & {\footnotesize (0.255)} & {\footnotesize (0.157)} \\ \cmidrule(r){2-10}
& & & \multirow{2}{*}{SI} & 0.065 & 0.067 & 0.063 & 0.050 & 0.079 & 0.062 \\
\multirow{2}{*}{(E2)} & \multirow{2}{*}{(F2)} & \multirow{2}{*}{3.744 {\footnotesize (1.004)}} & & {\footnotesize (0.246)} & {\footnotesize (0.250)} & {\footnotesize (0.244)} & {\footnotesize (0.218)} & {\footnotesize (0.270)} & {\footnotesize (0.153)} \\
& & & \multirow{2}{*}{Naive} & 1.000 & 1.000 & 1.000 & 1.000 & 1.000 & 0.999 \\
& & & & {\footnotesize (0.000)} & {\footnotesize (0.000)} & {\footnotesize (0.000)} & {\footnotesize (0.000)} & {\footnotesize (0.000)} & {\footnotesize (0.032)} \\ \cmidrule(r){2-10}
& & & \multirow{2}{*}{SI} & 0.045 & 0.066 & 0.056 & 0.053 & 0.050 & 0.042 \\
& \multirow{2}{*}{(F3)} & \multirow{2}{*}{1.916 {\footnotesize (1.191)}} & & {\footnotesize (0.207)} & {\footnotesize (0.248)} & {\footnotesize (0.230)} & {\footnotesize (0.224)} & {\footnotesize (0.219)} & {\footnotesize (0.145)} \\
& & & \multirow{2}{*}{Naive} & 1.000 & 1.000 & 1.000 & 1.000 & 1.000 & 0.886 \\
& & & & {\footnotesize (0.000)} & {\footnotesize (0.000)} & {\footnotesize (0.000)} & {\footnotesize (0.000)} & {\footnotesize (0.000)} & {\footnotesize (0.318)} \\
\bottomrule
\end{tabular}
}
\label{t:1}
\end{table}%

Table \ref{t:2} shows results for when (M2) is the causal effect. Here, the confounding variables followed a continuous uniform distribution, $\delta_n=(p/6)^{1/2}$ and $\lambda=5\sigma n^{-1/2}(\log p)^{1/2}$. In this case, all variables are active, so the closer the non-coverage probability is to 1, the better. The non-coverage probability of both methods is close to 1, but Naive does not keep the FCR at 0.05. On the other hand, SI keeps it at almost 0.05.

\begin{table}[t!]
\caption{Case in which $p=5$, $X_{ij}\sim{\rm U}(0,1)$ and (M2) is correct. The figures in parentheses are standard deviations. In all settings, SI removes a high percentage of 0's from the confidence intervals while appropriately controlling the FCR. }
\centering
{\small
\begin{tabular}{cccccccccc}
\toprule
$e(\bm{x})$ & $f(\bm{x})$ & $|\hat{M}|$ & & $\beta_1$ & $\beta_2$ & $\beta_3$ & $\beta_4$ & $\beta_5$ & FCR \\ \midrule
& & & \multirow{2}{*}{SI} & 1.000 & 0.998 & 0.998 & 0.999 & 0.999 & 0.054 \\
& \multirow{2}{*}{(F1)} & \multirow{2}{*}{4.987 {\footnotesize (0.113)}} & & {\footnotesize (0.000)} & {\footnotesize (0.045)} & {\footnotesize (0.045)} & {\footnotesize (0.032)} & {\footnotesize (0.032)} & {\footnotesize (0.099)} \\
& & & \multirow{2}{*}{Naive} & 1.000 & 1.000 & 1.000 & 1.000 & 1.000 & 0.417 \\
& & & & {\footnotesize (0.000)} & {\footnotesize (0.000)} & {\footnotesize (0.000)} & {\footnotesize (0.000)} & {\footnotesize (0.000)} & {\footnotesize (0.233)} \\ \cmidrule(r){2-10}
& & & \multirow{2}{*}{SI} & 1.000 & 0.994 & 0.995 & 0.992 & 0.990 & 0.067 \\
\multirow{2}{*}{(E1)} & \multirow{2}{*}{(F2)} & \multirow{2}{*}{4.460 {\footnotesize (0.754)}} & & {\footnotesize (0.000)} & {\footnotesize (0.076)} & {\footnotesize (0.068)} & {\footnotesize (0.089)} & {\footnotesize (0.101)} & {\footnotesize (0.127)} \\
& & & \multirow{2}{*}{Naive} & 1.000 & 1.000 & 1.000 & 1.000 & 1.000 & 0.737 \\
& & & & {\footnotesize (0.000)} & {\footnotesize (0.000)} & {\footnotesize (0.000)} & {\footnotesize (0.000)} & {\footnotesize (0.000)} & {\footnotesize (0.215)} \\ \cmidrule(r){2-10}
& & & \multirow{2}{*}{SI} & 1.000 & 0.998 & 0.991 & 0.999 & 0.997 & 0.064 \\
& \multirow{2}{*}{(F3)} & \multirow{2}{*}{4.888 {\footnotesize (0.334)}} & & {\footnotesize (0.000)} & {\footnotesize (0.045)} & {\footnotesize (0.096)} & {\footnotesize (0.032)} & {\footnotesize (0.055)} & {\footnotesize (0.114)} \\
& & & \multirow{2}{*}{Naive} & 1.000 & 1.000 & 1.000 & 1.000 & 1.000 & 0.591 \\
& & & & {\footnotesize (0.000)} & {\footnotesize (0.000)} & {\footnotesize (0.000)} & {\footnotesize (0.000)} & {\footnotesize (0.000)} & {\footnotesize (0.230)} \\
\midrule
& & & \multirow{2}{*}{SI} & 1.000 & 1.000 & 0.998 & 0.999 & 0.999 & 0.054 \\
& \multirow{2}{*}{(F1)} & \multirow{2}{*}{4.980 {\footnotesize (0.140)}} & & {\footnotesize (0.000)} & {\footnotesize (0.000)} & {\footnotesize (0.045)} & {\footnotesize (0.032)} & {\footnotesize (0.032)} & {\footnotesize (0.101)} \\
& & & \multirow{2}{*}{Naive} & 1.000 & 1.000 & 1.000 & 1.000 & 1.000 & 0.458 \\
& & & & {\footnotesize (0.000)} & {\footnotesize (0.000)} & {\footnotesize (0.000)} & {\footnotesize (0.000)} & {\footnotesize (0.000)} & {\footnotesize (0.227)} \\ \cmidrule(r){2-10}
& & & \multirow{2}{*}{SI} & 1.000 & 0.992 & 0.987 & 0.993 & 0.991 & 0.066 \\
\multirow{2}{*}{(E2)} & \multirow{2}{*}{(F2)} & \multirow{2}{*}{4.445 {\footnotesize (0.778)}} & & {\footnotesize (0.000)} & {\footnotesize (0.089)} & {\footnotesize (0.112)} & {\footnotesize (0.083)} & {\footnotesize (0.097)} & {\footnotesize (0.125)} \\
& & & \multirow{2}{*}{Naive} & 1.000 & 1.000 & 1.000 & 1.000 & 1.000 & 0.749 \\
& & & & {\footnotesize (0.000)} & {\footnotesize (0.000)} & {\footnotesize (0.000)} & {\footnotesize (0.000)} & {\footnotesize (0.000)} & {\footnotesize (0.221)} \\ \cmidrule(r){2-10}
& & & \multirow{2}{*}{SI} & 1.000 & 0.998 & 0.998 & 0.997 & 0.996 & 0.057 \\
& \multirow{2}{*}{(F3)} & \multirow{2}{*}{4.864 {\footnotesize (0.384)}} & & {\footnotesize (0.000)} & {\footnotesize (0.045)} & {\footnotesize (0.046)} & {\footnotesize (0.056)} & {\footnotesize (0.064)} & {\footnotesize (0.108)} \\
& & & \multirow{2}{*}{Naive} & 1.000 & 1.000 & 1.000 & 1.000 & 1.000 & 0.614 \\
& & & &{\footnotesize (0.000)} & {\footnotesize (0.000)} & {\footnotesize (0.000)} & {\footnotesize (0.000)} & {\footnotesize (0.000)} & {\footnotesize (0.228)} \\
\bottomrule
\end{tabular}
}
\label{t:2}
\end{table}%

\begin{figure}[t!]
\subfloat[Case of (E1) and (F3) in the setup of Table \ref{t:1}]{\includegraphics[width = 0.5\textwidth, bb = 0 0 504 360]{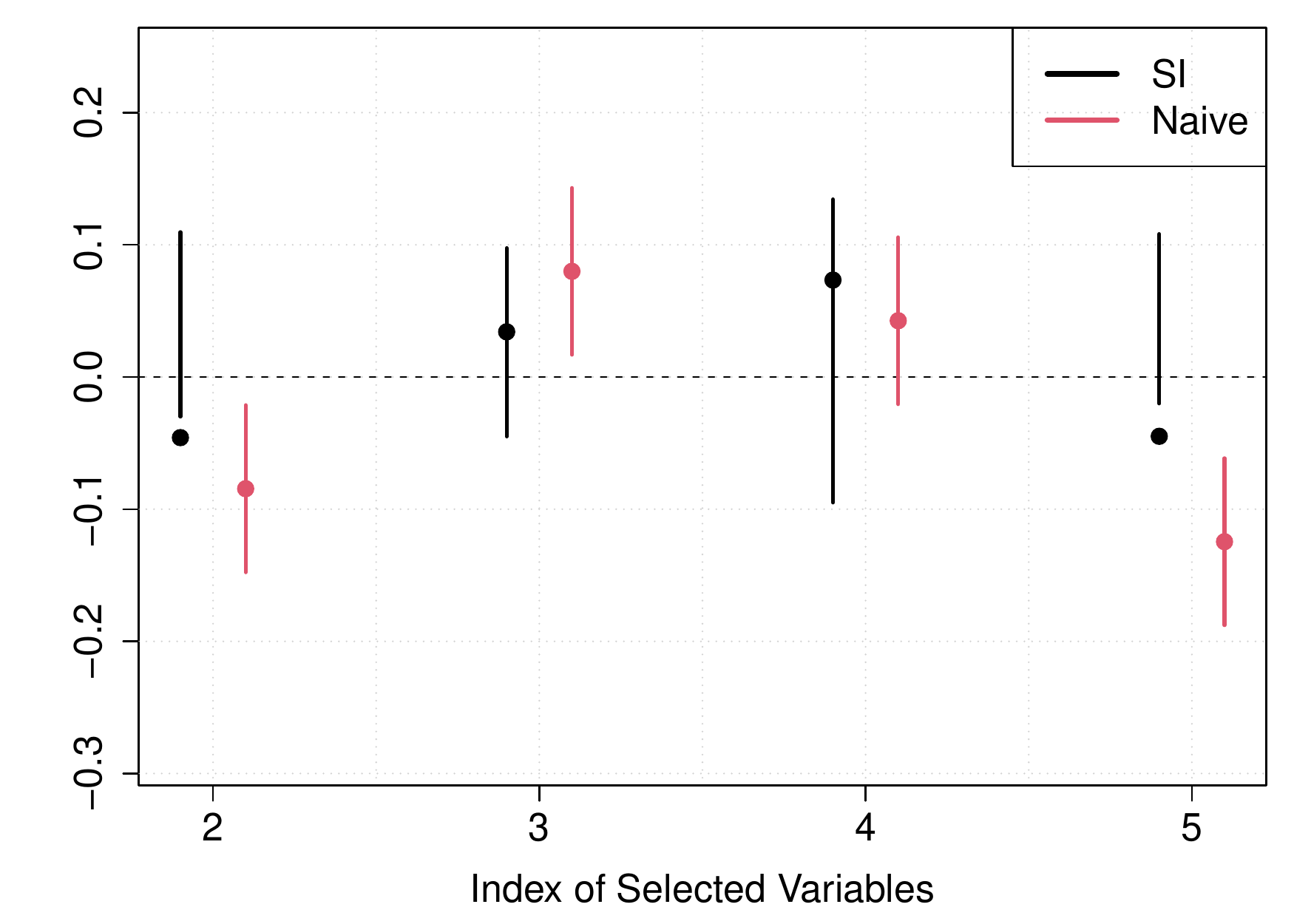}\label{f:1a}}
\subfloat[Case of (E2) and (F3) in the setup of Table \ref{t:1}]{\includegraphics[width = 0.5\textwidth, bb = 0 0 504 360]{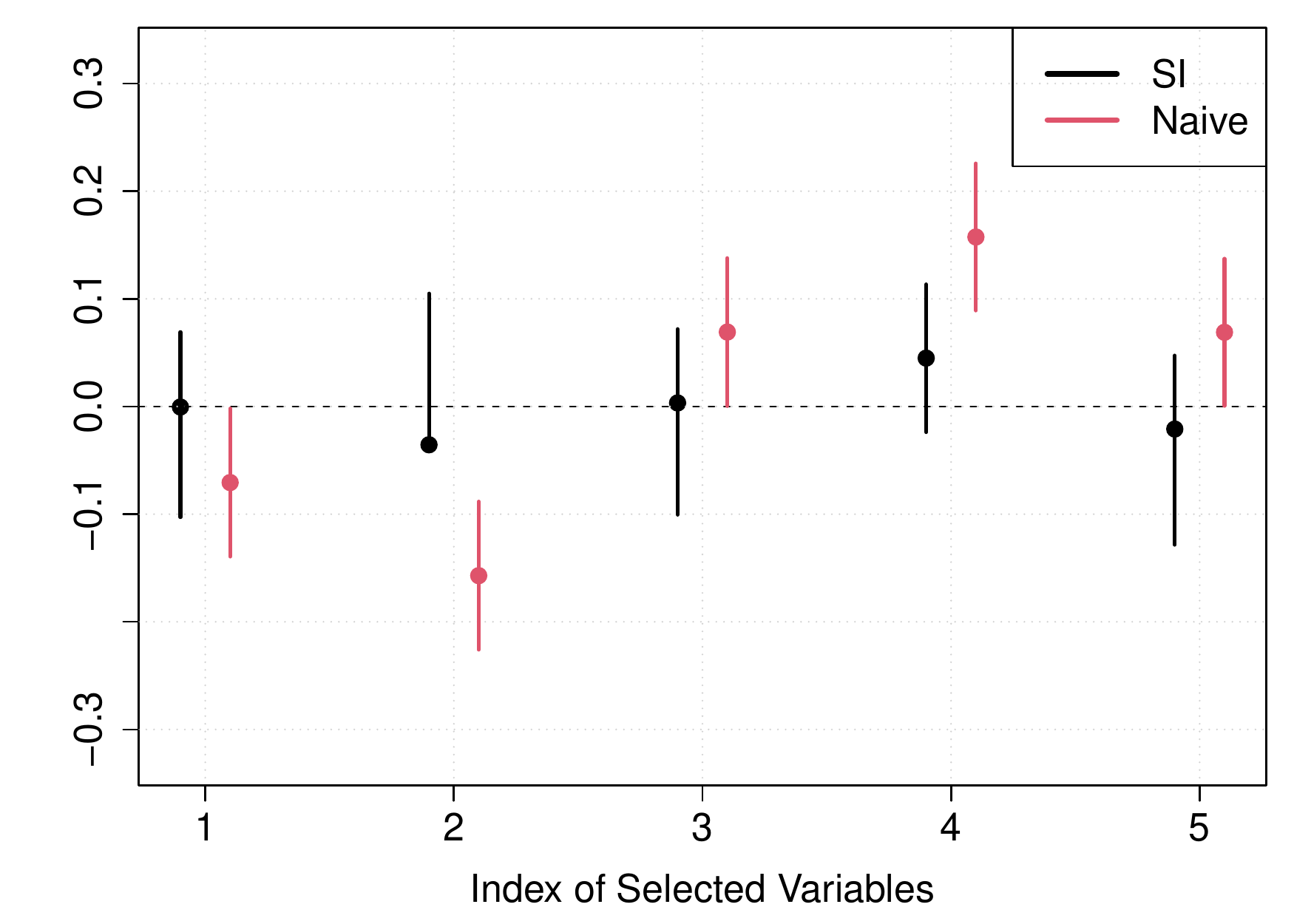}\label{f:1b}} \\
\subfloat[Case of (E1) and (F3) in the setup of Table \ref{t:2}]{\includegraphics[width = 0.5\textwidth, bb = 0 0 504 360]{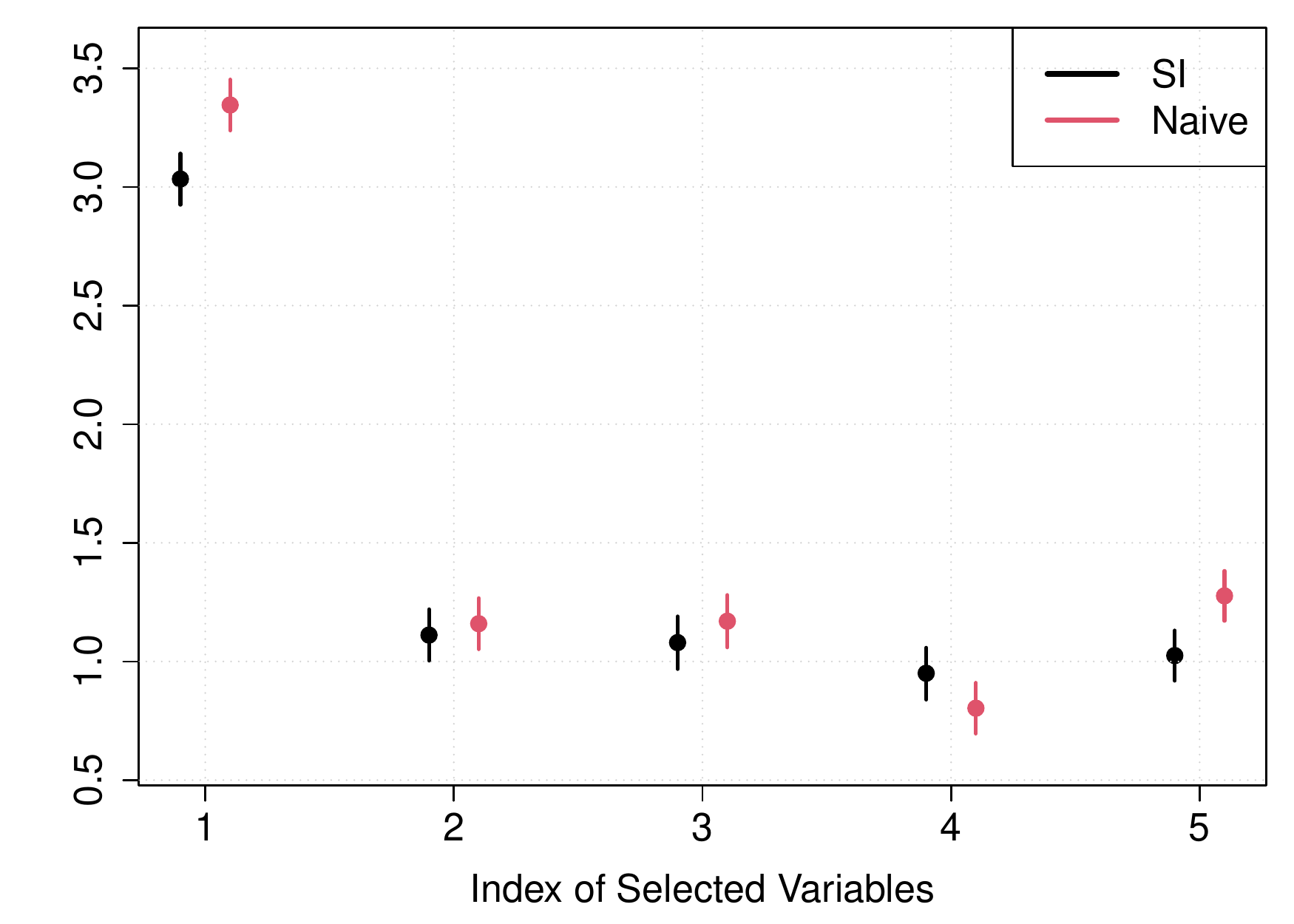}\label{f:1c}}
\subfloat[Case of (E2) and (F3) in the setup of Table \ref{t:2}]{\includegraphics[width = 0.5\textwidth, bb = 0 0 504 360]{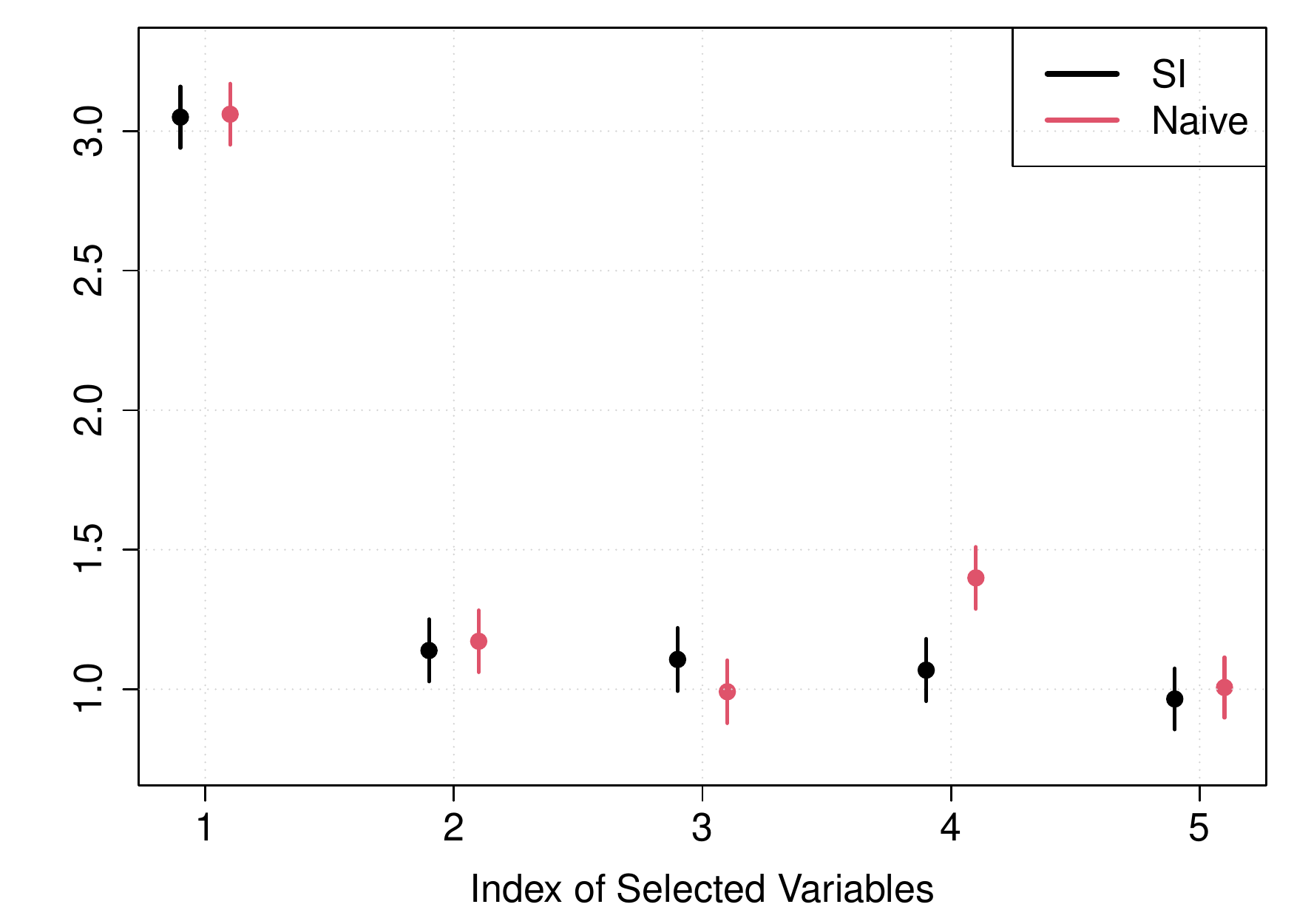}\label{f:1d}}
\caption{Comparison of confidence intervals for SI and Naive. The black (red) dots are the estimates for SI (Naive), and the solid lines are the confidence intervals with a confidence coefficient of 0.95. }
\label{f:1}
\end{figure}

Figure \ref{f:1} compares the confidence intervals obtained by SI and Naive. In all panels, the nuisance parameter is (F3). Figures \ref{f:1}\subref{f:1a} and \ref{f:1}\subref{f:1b} respectively use (E1) and (E2) as the propensity scores in the setting of Table \ref{t:1}. On the other hand, Figures \ref{f:1}\subref{f:1c} and \ref{f:1}\subref{f:1d} respectively use (E1) and (E2) as the propensity scores in the setting of Table \ref{t:2}. Note that the horizontal axis represents the variables selected by Lasso, and in Figure \ref{f:1}\subref{f:1a}, the variables $x_2, x_3, x_4, x_5$ are selected, while in the remaining figures, all variables are selected. Since (M1) is correct in Figures \ref{f:1}\subref{f:1a} and \ref{f:1}\subref{f:1b}, the confidence intervals should contain 0. In both panels, it can be seen that most of Naive's results do not contain 0, but those of SI do contain 0. In Figure \ref{f:1}\subref{f:1a}, the two confidence intervals for SI do not include the estimates, which may seem somewhat strange, but a similar phenomenon has been reported in \cite{LiuMT18}. Figure \ref{f:1}\subref{f:1c} shows that the confidence intervals for SI appropriately include the true values of the parameters except for $x_2$, but the confidence intervals for Naive do not include the true values for any variable.

\subsection{Case of $p=25$}
\label{sec4_2}
Table \ref{t:3} shows the results for when (M2) is the causal effect; i.e., the first 5 variables are active and the remaining 20 variables are inactive. In all settings, the tuning parameter of Lasso was set to $\lambda=2\sigma n^{-1/2}(\log p)^{1/2}$. The sequence $\delta_n$ used to define the neighborhood ${\cal N}_i^{(h)}$ was set to $(p/2)^{1/2}$ for $X_{ij}\sim{\rm Ber}(1/2)$ and $(p/6)^{1/2}$ for $X_{ij}\sim{\rm U}(0,1)$. As in the case of $p=5$, we can see that SI appropriately controls the FCR, keeping it at almost 0.05 in all settings. In addition, since TP is close to 5, the confidence intervals for the active variables are appropriately away from 0, and FP is sufficiently small. On the other hand, the FCR for Naive is well above 0.05, suggesting that reasonable confidence intervals were not obtained. The values of TP, FP, and $|\hat{M}|$ indicate that almost all of the confidence intervals for the variables selected by Lasso are away from 0.

\begin{table}[t!]
\caption{Cases in which $p=25$ and (M2) is correct. The figures in parentheses are standard deviations. In all settings, SI controls the FCR appropriately, but the inference of Naive is invalid. }
\centering
{\small
\begin{tabular}{ccccccccccc}
\toprule
& & \multicolumn{4}{c}{$X_{ij}\sim{\rm Ber}(1/2)$} & \multicolumn{4}{c}{$X_{ij}\sim{\rm U}(0,1)$} \\ \cmidrule(r){4-7} \cmidrule(r){8-11}
$e(\bm{x})$ & $f(\bm{x})$ & & $|\hat{M}|$ & TP & FP & FCR & $|\hat{M}|$ & TP & FP & FCR \\ \midrule
& & \multirow{2}{*}{SI} & & 4.991 & 0.323 & 0.049 & & 5.000 &0.002 & 0.051 \\
& \multirow{2}{*}{(F1)} & &11.775 & {\footnotesize (0.114)} & {\footnotesize (0.592)} & {\footnotesize (0.068)} & 5.143 & {\footnotesize (0.000)} & {\footnotesize (0.045)} & {\footnotesize (0.098)} \\
& & \multirow{2}{*}{Naive} & {\footnotesize (2.154)} & 5.000 & 6.775 & 0.830 & {\footnotesize (0.380)} & 5.000 & 0.143 & 0.436 \\
& & & & {\footnotesize (0.000)} & {\footnotesize (2.154)} & {\footnotesize (0.106)} & & {\footnotesize (0.000)} & {\footnotesize (0.380)} & {\footnotesize (0.234)} \\ \cmidrule(r){2-11}
& & \multirow{2}{*}{SI} & & 4.972 & 0.879 & 0.058 & & 4.827 & 0.392 & 0.058 \\
\multirow{2}{*}{(E1)} & \multirow{2}{*}{(F2)} & & 19.804 & {\footnotesize (0.213)} & {\footnotesize (1.213)} & {\footnotesize (0.074)} & 11.399 & {\footnotesize (0.423)} & {\footnotesize (0.647)} & {\footnotesize (0.079)} \\
& & \multirow{2}{*}{Naive} & {\footnotesize (1.979)} & 4.982 & 14.822 & 0.965 & {\footnotesize (2.023)} & 4.852 & 6.547 & 0.893 \\
& & & & {\footnotesize (0.133)} & {\footnotesize (1.979)} & {\footnotesize (0.040)} & & {\footnotesize (0.385)} & {\footnotesize (1.997)} & {\footnotesize (0.093)} \\ \cmidrule(r){2-11}
& & \multirow{2}{*}{SI} & & 4.990 & 0.573 & 0.053 & & 4.992 & 0.076 & 0.053 \\
& \multirow{2}{*}{(F3)} & & 15.960 & {\footnotesize (0.148)} & {\footnotesize (0.847)} & {\footnotesize (0.065)} & 6.676 & {\footnotesize (0.126)} & {\footnotesize (0.273)} & {\footnotesize (0.089)} \\
& & \multirow{2}{*}{Naive} & {\footnotesize (2.198)} & 5.000 & 10.960 & 0.921 & {\footnotesize (1.230)} & 4.998 & 1.678 & 0.690 \\
& & & & {\footnotesize (0.000)} & {\footnotesize (2.198)} & {\footnotesize (0.066)} & & {\footnotesize (0.045)} & {\footnotesize (1.230)} & {\footnotesize (0.187)} \\
\midrule
& & \multirow{2}{*}{SI} & & 4.979 & 0.405 & 0.048 & & 4.994 & 0.013 & 0.053 \\
& \multirow{2}{*}{(F1)} & & 13.841 & {\footnotesize (0.211)} & {\footnotesize (0.729)} & {\footnotesize (0.080)} & 5.287 & {\footnotesize (0.089)} & {\footnotesize (0.113)} & {\footnotesize (0.098)} \\
& & \multirow{2}{*}{Naive} & {\footnotesize (2.322)} & 5.000 & 8.841 & 0.874 & {\footnotesize (0.539)} & 5.000 & 0.287 & 0.491 \\
& & & & {\footnotesize (0.000)} & {\footnotesize (2.322)} & {\footnotesize (0.089)} & & {\footnotesize (0.000)} & {\footnotesize (0.539)} & {\footnotesize (0.229)} \\ \cmidrule(r){2-11}
& & \multirow{2}{*}{SI} & & 4.938 & 0.834 & 0.053 & & 4.780 & 0.391 & 0.054 \\
\multirow{2}{*}{(E2)} & \multirow{2}{*}{(F2)} & & 20.380 & {\footnotesize (0.280)} & {\footnotesize (0.935)} & {\footnotesize (0.056)} & 11.844 & {\footnotesize (0.498)} & {\footnotesize (0.673)} & {\footnotesize (0.077)} \\
& & \multirow{2}{*}{Naive} & {\footnotesize (1.906)} & 4.954 & 15.425 & 0.968 & {\footnotesize (2.121)} & 4.818 & 7.026 & 0.904 \\
& & & & {\footnotesize (0.214)} & {\footnotesize (1.887)} & {\footnotesize (0.038)} & & {\footnotesize (0.435)} & {\footnotesize (2.085)} & {\footnotesize (0.084)} \\ \cmidrule(r){2-11}
& & \multirow{2}{*}{SI} & & 4.963 & 0.647 & 0.052 & & 4.974 & 0.101 & 0.054 \\
& \multirow{2}{*}{(F3)} & & 17.238 & {\footnotesize (0.309)} & {\footnotesize (1.028)} & {\footnotesize (0.075)} & 7.050 & {\footnotesize (0.222)} & {\footnotesize (0.359)} & {\footnotesize (0.099)} \\
& & \multirow{2}{*}{Naive} & {\footnotesize (2.244)} & 5.000 & 12.238 & 0.933 & {\footnotesize (1.361)} & 4.997 & 2.053 & 0.724 \\
& & & & {\footnotesize (0.000)} & {\footnotesize (2.244)} & {\footnotesize (0.059)} & & {\footnotesize (0.055)} & {\footnotesize (1.359)} & {\footnotesize (0.174)} \\
\bottomrule
\end{tabular}
}
\label{t:3}
\end{table}%

\begin{figure}[t!]
\subfloat[Case of $X_{ij}\sim{\rm U}(0, 1)$]{\includegraphics[width = 0.5\textwidth, bb = 0 0 504 360]{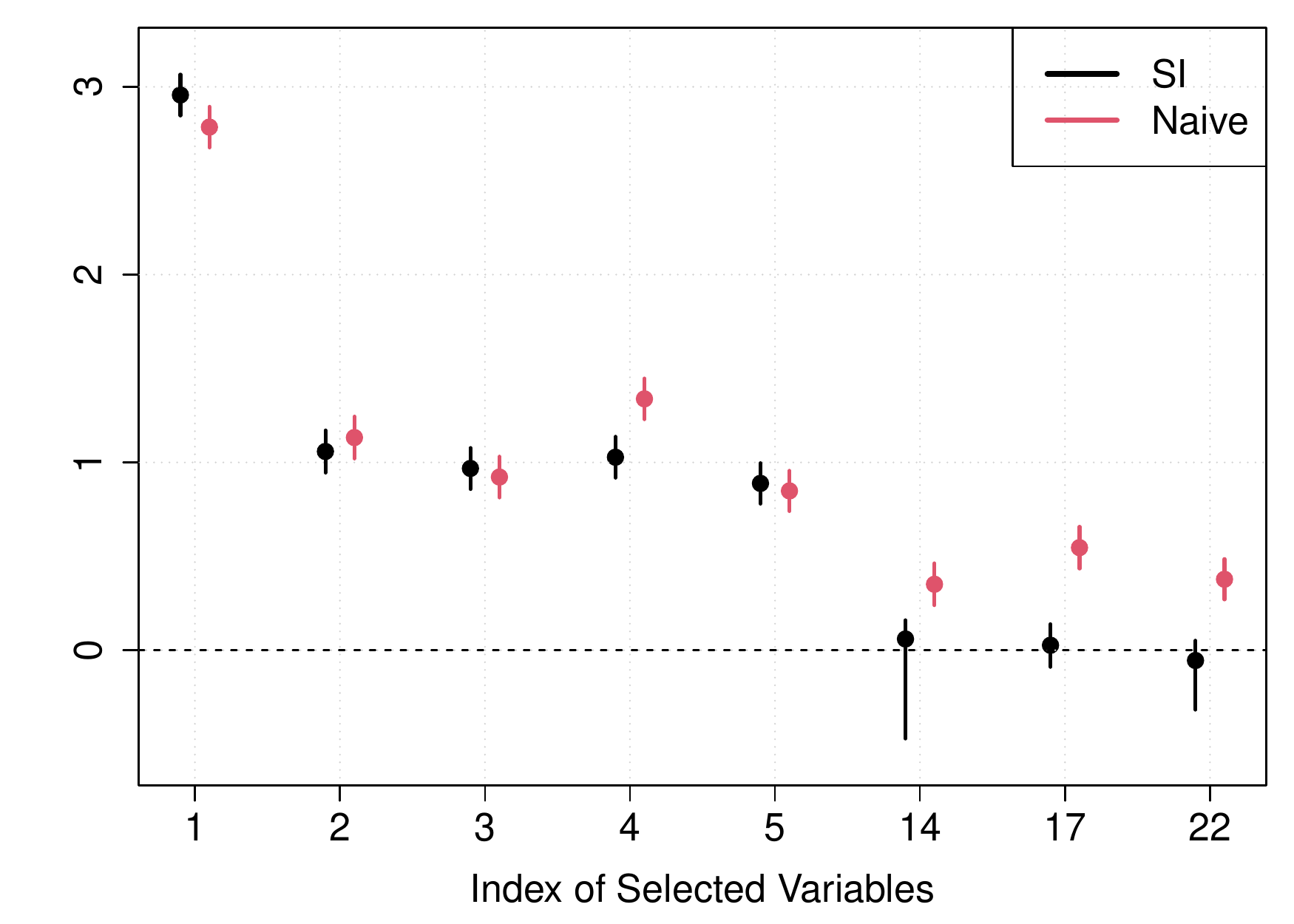}}
\subfloat[Case of $X_{ij}\sim{\rm Ber}(1/2)$]{\includegraphics[width = 0.5\textwidth, bb = 0 0 504 360]{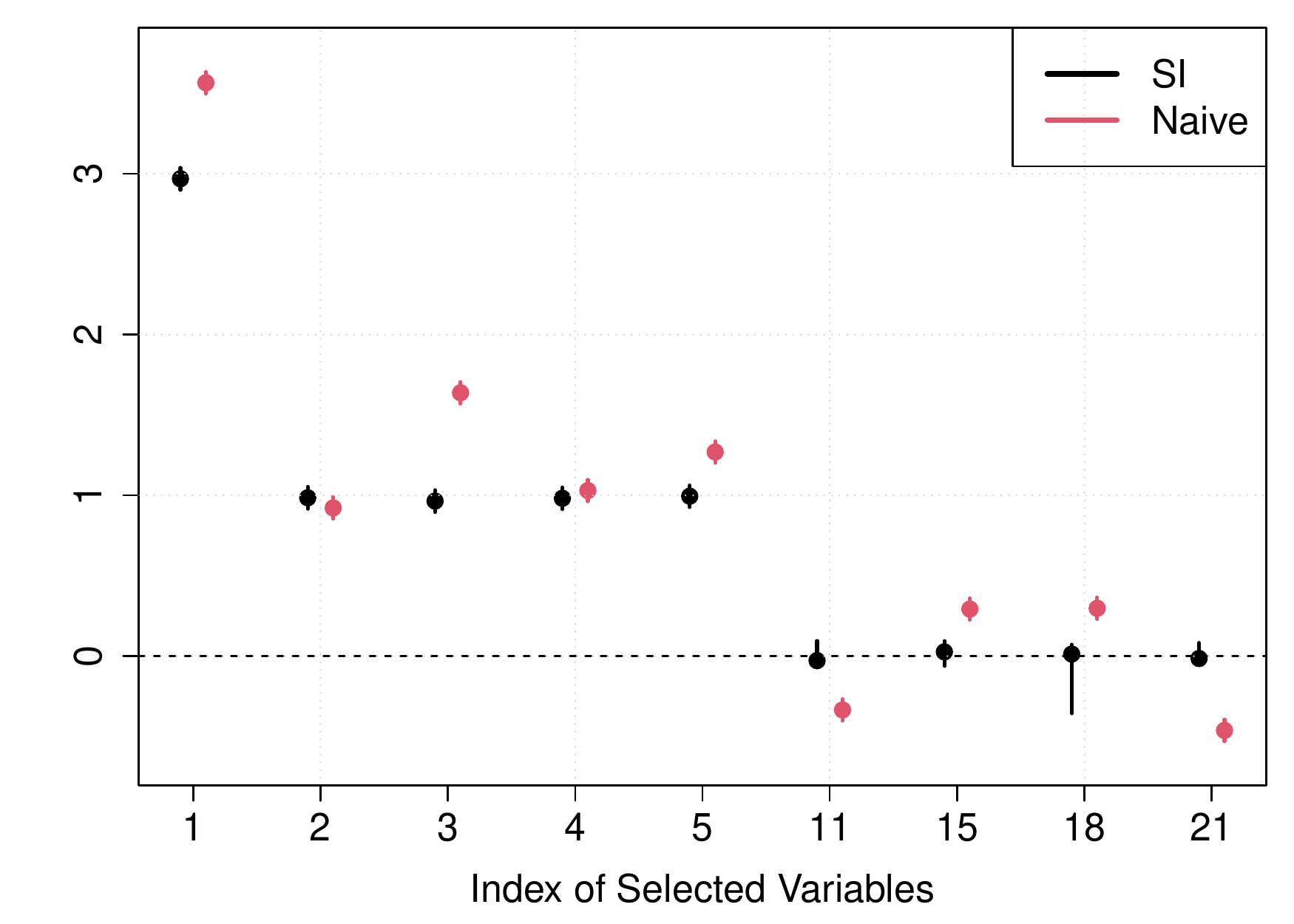}}
\caption{Comparison of confidence intervals for SI and Naive for cases using (F3), (E2), and (M2) as models. The black (red) dots are the estimates for SI (Naive), and the solid lines are the confidence intervals with a confidence coefficient of 0.95. }
\label{f:2}
\end{figure}

Figure \ref{f:2} compares the confidence intervals for SI and Naive with a confidence coefficient of 0.95 where the nuisance parameters and propensity scores are (F3) and (E2). As explained in Table \ref{t:3}, Naive has extremely short confidence intervals due to loss of the validity of the inference, and the confidence interval does not include 0 even when the true value is 0. On the other hand, the confidence intervals for SI contain a high percentage of true values.

\section{Data analysis}
\label{sec5}
 In practice, the error variance $\sigma^2$ and propensity score $e(\bm{x})$ must be estimated in some way. As described in Sections \ref{sec6_1} and \ref{sec6_3}, appropriate estimators can be substituted for them. Thereafter, the error variance estimate $\hat{\sigma}^2$ can be calculated using the method in Section \ref{sec6_1} as $\delta_n=(p/2)^{1/2}$, and the propensity score estimate $\hat{e}(\bm{x})$ can be calculated from a logistic regression model. Using the estimates, the tuning parameter of Lasso was set to $\lambda=\hat{\sigma}n^{-1/2}(\log p)^{1/2}$. The following subsections describe the application of the proposed method to two real datasets.

\subsection{Application to the lalonde dataset}
\label{sec5_1}
The lalonde dataset is treated in \cite{Lal86} and is included in the R package Matching. We set the group that took the U.S. job training program in 1976 as $t=1$ and the group that did not take the program as $t=0$. We predicted the difference in annual income in 1978 with training as a causal effect. The confounding variables were age (age), years of education (educ), black or not (black), Hispanic or not (hisp), married or not (married), high school graduate or higher (nodegr), income in 1974 (re74), income in 1975 (re75), zero income in 1974 or not (u74), zero income in 1975 or not (u75), i.e., $p=10$, and the outcome variable was income in 1978 (re78). The sample size was $n=445$.

When Lasso was used, 5 variables were selected: age, educ, re74, re75 and u74. For these coefficients, we compared the confidence intervals with a confidence coefficient of 0.95 given by the proposed method (SI) and a method ignoring the influence of model selection (Naive). As shown in Table \ref{t:4}, for all variables, the confidence interval for Naive is shorter than that for SI. This is because, as mentioned in the previous section, Naive ignores that the selected variable is likely to be significant. In both methods, the confidence intervals for educ and u74 do not contain 0, indicating that they are significant, while the confidence intervals for age and re75 contain 0, indicating that they are not significant. On the other hand, SI and Naive give different results for re74. Since u74 is a variable that is 0 if there is income in 1974 and 1 if there is not, if u74 is 1, the corresponding re74 is 0. Therefore, the SI result can be interpreted as that the causal effect is affected by whether there was income or not in 1974, rather than by the size of the income in 1974 itself.

\begin{table}[t!]
\caption{ Confidence intervals with a confidence coefficient of 0.95 for the 5 variables obtained by applying Lasso to the lalonde dataset. }
\centering
{\small
\begin{tabular}{crrrrrr}
\toprule
& \multicolumn{3}{c}{SI} & \multicolumn{3}{c}{Naive} \\ \cmidrule(r){2-4}\cmidrule(r){5-7}
Selected variable & Estimates & lower & upper & Estimates & lower & upper \\ \midrule
age & $49.353$ & $-164.745$ & $165.970$ & $51.345$ & $-66.302$ & $168.991$ \\
educ & $721.188$ & $228.418$ & $1196.124$ & $862.309$ & $390.500$ & $1334.117$ \\
re74 & $0.240$ & $-0.173$ & $0.479$ & $0.236$ & $0.019$ & $0.454$ \\
re75 & $0.275$ & $-0.341$ & $0.645$ & $0.293$ & $-0.040$ & $0.625$ \\
u74 & $6459.088$ & $3709.181$ & $8956.314$ & $6480.235$ & $3997.591$ & $8962.880$ \\
\bottomrule
\end{tabular}
}
\label{t:4}
\end{table}%

\subsection{Application to E-commerce dataset}
\label{sec5_2}
The MineThatData dataset is published by \cite{Hil08} and contains 12 variables such as purchase amount and district classification code for 64,000 customers. The objective is to predict the purchase price of a product from the type of email delivered (male, female, or unsent). The sample size of 64,000 was too large computational load for our resources, so we set the assignment variable to that of emails delivered to males ($t=1$) and unsent emails ($t=0$). Moreover, we set the sample size to $n=6,382$ and the district code (zip\_code) to Rural. The confounding variables were the number of months since the last purchase (recency), the amount of purchases made in the last year (history), whether the user purchased men's products in the last year (mens), whether the user purchased women's products in the last year (womens), whether the user became a new user in the last 12 months (newbie), whether the user received the email within two weeks of the delivery (newbie), whether the user visited the site within two weeks after the mail was delivered (visit), and whether the user made a purchase within two weeks after the mail was delivered (conversion), i.e., $p=7$. The outcome variable was the purchase amount when the user made a purchase (spend).

After the preprocessing described above, the error variance and propensity scores were estimated, and Lasso selected four variables: recency, history, newbie and conversion. Table \ref{t:5} compares the confidence intervals with a confidence coefficient of 0.95 for these coefficients. In the case of Naive, all four variables are significant, while in the case of SI, only history and conversion are significant. A natural interpretation of the SI result is that it is important for a customer to make a large number of purchases within two weeks of receiving an email in order to predict the purchase amount. In the case of Naive, history is negative and significant, indicating that a large purchase amount over the preceding year has a negative effect on the prediction of the purchase amount, which is rather unnatural. On the other hand, in SI, although its estimate is negative, its confidence interval is positive; i.e., it is a natural interval estimation.

\begin{table}[t!]
\caption{ Confidence intervals with a confidence coefficient of 0.95 for the 4 variables obtained by applying Lasso to the E-commerce dataset. }
\centering
{\small
\begin{tabular}{crrrrrr}
\toprule
& \multicolumn{3}{c}{SI} & \multicolumn{3}{c}{Naive} \\ \cmidrule(r){2-4}\cmidrule(r){5-7}
Selected variable & Estimates & lower & upper & Estimates & lower & upper \\ \midrule
recency & $-0.094$ & $-0.059$ & $0.225$ & $-0.143$ & $-0.274$ & $-0.011$ \\
history & $-0.002$ & $0.000$ & $0.004$ & $-0.003$ & $-0.005$ & $-0.001$ \\
newbie & $0.704$ & $-0.640$ & $1.631$ & $1.419$ & $0.494$ & $2.343$ \\
conversion & $13.378$ & $8.661$ & $17.857$ & $20.191$ & $15.730$ & $24.651$ \\
\bottomrule
\end{tabular}
}
\label{t:5}
\end{table}%

\section{Extension}
\label{sec6}
\subsection{Case of unknown variance}
\label{sec6_1}
Up to this point, we have treated the variance $\sigma^2$ of $\epsilon^{(h)}$ in \eqref{model1} as known. However, it is usually unknown in practical applications. As such, it would seem that we only have to substitute a consistent estimator, but since the model contains $f(\cdot)$ and we are trying to develop a method that avoids its identification, its determination is more difficult than usual. Here, let us choose an appropriate sequence of real numbers $\{\delta_n\}$ that converges to $0$, define ${\mathcal N}_{i}^{\dagger}=\{l\neq i:\ \|\bm{X}_l-\bm{X}_i\|_2<\delta_n,\ \bm{T}_l=\bm{T}_i\}$ using $\bm{T}_i=(T_i^{(1)},\ldots,T_i^{(H)})$, and use
\begin{align*}
\frac{1}{n}\sum_{i=1}^n\frac{|{\mathcal N}_{i}^{\dagger}|}{1+|{\mathcal N}_{i}^{\dagger}|} \bigg( Y_i-\frac{1}{|{\mathcal N}_{i}^{\dagger}|} \sum_{l\in{\mathcal N}_{i}^{\dagger}} Y_l \bigg)^2.
\end{align*}
If the set of possible values of $\bm{X}_i$ is finite, or if the set is compact and $f(\cdot)$ is continuous, this estimator is obviously a consistent estimator of $\sigma^2$.

\subsection{Generalization of error structure, model selection method, and causal effect}
\label{sec6_2}
In Section \ref{sec3}, we developed selective inference for basic causal inference models. There is no difficulty in extending it to the case of certain error structures, model selection methods, and causal effects. Here, we assume that, in \eqref{model1}, the variance of $\bm{\epsilon}^{(h)}\ (\in\mathbb{R}^r)$ is $\bm{\Sigma}$ in the setup of multivariate regression, conduct a model selection by using an elastic net (\citealt{ZouHas05}) as is done in \cite{LeeSST16}, and consider the causal effect $\E(\sum_{h=1}^Hc^{(h)}Y^{(h)}\mid T^{(1)}=1)$ in the group receiving the first treatment as an estimand. Note that this causal effect is the average treatment effect on the treated or the average treatment effect on the untreated when $H=2$ (\citealt{Imb04}). Below, we assume the strongly ignorable treatment assignment condition that assures a valid estimation. The variance of $\tilde{\bm{\epsilon}}^{(h)}\ (\in\mathbb{R}^{nr})$ in \eqref{model2} is $\tilde{\bm{\Sigma}}\otimes\bm{I}_n$, which we will denote as $\tilde{\bm{\Sigma}}$. Let $\lambda_1$ and $\lambda_2$ be the tuning parameters. We will use
\begin{align*}
\hat{\bm{\beta}} = \argmin_{\bm{\beta}} \bigg[ \frac{1}{2} \{ \tilde{\bm{W}} ( \tilde{\bm{T}},\tilde{\bm{X}} ) \tilde{\bm{Y}} - \tilde{\bm{X}} \bm{\beta} \}' \tilde{\bm{\Sigma}}^{-1} \{ \tilde{\bm{W}} ( \tilde{\bm{T}},\tilde{\bm{X}} ) \tilde{\bm{Y}} - \tilde{\bm{X}} \bm{\beta} \} + \lambda_1 \| \bm{\beta} \|_1 + \frac{\lambda_2}{2} \| \bm{\beta} \|_2^2 \bigg]
\end{align*}
instead of \eqref{IPWlasso}. Here, $\tilde{\bm{W}}(\tilde{\bm{T}},\tilde{\bm{X}})$ is $\sum_{h=1}^Hc^{(h)}\tilde{\bm{W}}^{(h)}(\tilde{\bm{T}}^{(h)},\tilde{\bm{X}})$, and $\tilde{\bm{W}}^{(h)}(\tilde{\bm{T}}^{(h)},\tilde{\bm{X}})\equiv{\rm diag}[T_1^{(h)}\allowbreak e^{(1)}(\bm{X}_1)/\{e^{(h)}(\bm{X}_1)\E(T_1^{(1)})\},\ldots,T_n^{(h)}e^{(1)}(\bm{X}_n)/\{e^{(h)}(\bm{X}_n)\E(T_n^{(1)})\}]$ is the weight matrix for estimating the causal effect described above.

In this setup, the coefficients in \eqref{condunion} are given by
\begin{align*}
\bm{A}(\bm{M},\bm{s}) = \frac{1}{\lambda_1} \left( \begin{array}{c} \tilde{\bm{X}}_{\bm{M}^{\rm c}}' \tilde{\bm{\Sigma}}^{-1} \{ \bm{I}_n-\tilde{\bm{X}}_{\bm{M}} ( \tilde{\bm{X}}_{\bm{M}}' \tilde{\bm{\Sigma}}^{-1} \tilde{\bm{X}}_{\bm{M}} + \lambda_2\bm{I}_{\bm{M}} )^{-1} \tilde{\bm{X}}_{\bm{M}}' \tilde{\bm{\Sigma}}^{-1} \} \\ -\tilde{\bm{X}}_{\bm{M}^{\rm c}}' \tilde{\bm{\Sigma}}^{-1} \{ \bm{I}_n-\tilde{\bm{X}}_{\bm{M}} ( \tilde{\bm{X}}_{\bm{M}}' \tilde{\bm{\Sigma}}^{-1} \tilde{\bm{X}}_{\bm{M}} + \lambda_2\bm{I}_{\bm{M}} )^{-1} \tilde{\bm{X}}_{\bm{M}}' \tilde{\bm{\Sigma}}^{-1} \} \\ -{\rm diag}(\bm{s}) ( \tilde{\bm{X}}_{\bm{M}}' \tilde{\bm{\Sigma}}^{-1} \tilde{\bm{X}}_{\bm{M}} + \lambda_2\bm{I}_{\bm{M}} )^{-1} \tilde{\bm{X}}_{\bm{M}}' \tilde{\bm{\Sigma}}^{-1} \end{array} \right)
\end{align*}
and
\begin{align*}
\bm{b}(\bm{M},\bm{s}) = \left( \begin{array}{c} \bm{1}_{|\bm{M}^{\rm c}|}-\tilde{\bm{X}}_{\bm{M}^{\rm c}}' \tilde{\bm{\Sigma}}^{-1} \tilde{\bm{X}}_{\bm{M}} ( \tilde{\bm{X}}_{\bm{M}}' \tilde{\bm{\Sigma}}^{-1} \tilde{\bm{X}}_{\bm{M}} + \lambda_2\bm{I}_{\bm{M}} )^{-1} \bm{s} \\ \bm{1}_{|\bm{M}^{\rm c}|}+\tilde{\bm{X}}_{\bm{M}^{\rm c}}' \tilde{\bm{\Sigma}}^{-1} \tilde{\bm{X}}_{\bm{M}} ( \tilde{\bm{X}}_{\bm{M}}' \tilde{\bm{\Sigma}}^{-1} \tilde{\bm{X}}_{\bm{M}} + \lambda_2\bm{I}_{\bm{M}} )^{-1} \bm{s} \\ -{\rm diag}(\bm{s}) ( \tilde{\bm{X}}_{\bm{M}}' \tilde{\bm{\Sigma}}^{-1} \tilde{\bm{X}}_{\bm{M}} + \lambda_2\bm{I}_{\bm{M}} )^{-1} \bm{s} \end{array} \right).
\end{align*}
Then, by letting $\tilde{\bm{E}}(\tilde{\bm{X}})\equiv{\rm diag}\{e^{(1)}(\bm{X}_1),\ldots,e^{(1)}(\bm{X}_n)\}$, $\tilde{\bm{K}}^{(h)}(\tilde{\bm{X}})\equiv{\rm diag}[e^{(1)}(\bm{X}_1)\allowbreak\{1/e^{(h)}(\bm{X}_1)-1\}/|{\mathcal N}_1^{(h)}|,\ldots,e^{(1)}(\bm{X}_n)\{1/e^{(h)}(\bm{X}_n)-1\}/|{\mathcal N}_n^{(h)}|]$,
\begin{align*}
& \tilde{\bm{\eta}}_j = \tilde{\bm{\Sigma}}^{-1} \tilde{\bm{X}}_{\bm{M}} (\tilde{\bm{X}}_{\bm{M}}' \tilde{\bm{\Sigma}}^{-1} \tilde{\bm{X}}_{\bm{M}})^{-1} \bm{e}_j,
\\
& \kappa_j^{\bm{M}} ( \tilde{\bm{T}},\tilde{\bm{X}} ) = \bm{e}_j' ( \tilde{\bm{X}}_{\bm{M}}' \tilde{\bm{\Sigma}}^{-1} \tilde{\bm{X}}_{\bm{M}} )^{-1} \tilde{\bm{X}}_{\bm{M}}' \tilde{\bm{\Sigma}}^{-1} \sum_{h=1}^H c^{(h)} \tilde{\bm{W}}^{(h)} ( \tilde{\bm{T}}^{(h)},\tilde{\bm{X}} ) \{ \tilde{\bm{\mu}}^{(h)}( \tilde{\bm{X}} ) + \tilde{\bm{f}} ( \tilde{\bm{X}} ) \},
\\
& \zeta_j^{\bm{M}} ( \tilde{\bm{T}},\tilde{\bm{X}} ) = \bm{e}_j' ( \tilde{\bm{X}}_{\bm{M}}' \tilde{\bm{\Sigma}}^{-1} \tilde{\bm{X}}_{\bm{M}} )^{-1} \tilde{\bm{X}}_{\bm{M}}' \tilde{\bm{\Sigma}}^{-1} \tilde{\bm{W}} ( \tilde{\bm{T}},\tilde{\bm{X}} ) \tilde{\bm{\Sigma}} \tilde{\bm{W}} ( \tilde{\bm{T}},\tilde{\bm{X}} ) \tilde{\bm{\Sigma}}^{-1} \tilde{\bm{X}}_{\bm{M}} ( \tilde{\bm{X}}_{\bm{M}}' \tilde{\bm{\Sigma}}^{-1} \tilde{\bm{X}}_{\bm{M}} )^{-1} \bm{e}_j, 
\\
& \tau_j^{\bm{M}} ( \tilde{\bm{Y}}^{\dagger},\tilde{\bm{T}},\tilde{\bm{X}} ) = \bm{e}_j' ( \tilde{\bm{X}}_{\bm{M}}' \tilde{\bm{\Sigma}}^{-1} \tilde{\bm{X}}_{\bm{M}} )^{-1} \tilde{\bm{X}}_{\bm{M}}' \tilde{\bm{\Sigma}}^{-1} \sum_{h=1}^H c^{(h)} \{ \tilde{\bm{W}}^{(h)} ( \tilde{\bm{T}}^{(h)},\tilde{\bm{X}}) - \tilde{\bm{E}}(\tilde{\bm{X}})\otimes\bm{I}_{r} \} \tilde{\bm{Y}}^{(h)\dagger}
\end{align*}
and
\begin{align*}
\rho_j^{\bm{M}} ( \tilde{\bm{T}},\tilde{\bm{X}} ) = \zeta_j^{\bm{M}} ( \tilde{\bm{T}},\tilde{\bm{X}} ) + \tilde{\bm{\eta}}_j' \sum_{h=1}^H c^{(h)2} \{ \tilde{\bm{K}}^{(h)}(\tilde{\bm{X}}) \otimes \bm{\Sigma} \} \tilde{\bm{\eta}}_j, 
\end{align*}
it can be shown that Theorem \ref{th1}, Lemma \ref{th2} and Theorem \ref{th3} hold.

\subsection{Case of unknown propensity score}
\label{sec6_3}
In Section \ref{sec3}, we treated the propensity scores as known, but in reality, they are often unknown and are usually estimated by performing some kind of discriminant analysis. We can formally use the results of Section \ref{sec3} as if the estimated propensity scores were known, but for the sake of a more rigorous methodology, we will take the effect of the estimation into account. We will regress $\tilde{\bm{x}}$ on $\tilde{\bm{t}}$ with some function and write $\hat{e}^{(h)}(\bm{X}_i)$ as the estimator of the propensity score $e^{(h)}(\bm{X}_i)=\P(T_i^{(h)}=1\mid\bm{X}_i)$. Here, we assume that there is no misspecification in the regression model of the propensity scores and that $\hat{e}^{(h)}(\bm{X}_i)-e^{(h)}(\bm{X}_i)=\OP(n^{-1/2})$. Note that the results in this section do not depend on the representation of this regression model.

In $\tilde{\bm{\bm{W}}}(\tilde{\bm{\bm{T}}},\tilde{\bm{\bm{X}}})$ which appears in Section \ref{sec3}, we replace $e^{(h)}(\bm{X}_i)$ with $\hat{e}^{(h)}(\bm{X}_i)$ and write it as $\hat{\tilde{\bm{W}}}(\tilde{\bm{T}},\tilde{\bm{X}})$. Note that under the conditioning of \eqref{condi1}, $\hat{e}^{(h)}(\bm{X}_i)$ is a random variable, but under the conditioning of \eqref{condi2}, which is unique to this paper, $\hat{e}^{(h)}(\bm{X}_i)$ is a non-random variable. This does not require using the asymptotic variance of the estimator when the propensity score is unknown, which has an interesting paradox of being smaller than when it is known (\citealt{HirIR03}). Then, Theorem \ref{th1} and Lemma \ref{th2} still hold. The different consideration is needed in the derivation of Theorem \ref{th3}, where we evaluate the variance of $\tilde{\bm{\eta}}_j' \hat{\tilde{\bm{W}}} ( \tilde{\bm{T}},\tilde{\bm{X}} ) \tilde{\bm{Y}}-\tau_j^{\bm{M}} ( \tilde{\bm{Y}}^*,\tilde{\bm{T}},\tilde{\bm{X}} )$. Because of the fluctuations in $\hat{e}^{(h)}(\bm{X}_i)$, the evaluation of the matrix in \eqref{CondE1} is modified to include $\E(\tilde{\bm{\epsilon}}^{(h)*} \tilde{\bm{\epsilon}}^{(k)*}{}')\times\OP(n^{-1})$. As a result, the conditional expectation of the second terms still has the form \eqref{CondE3}. As well, the evaluation of the matrix in \eqref{CondE2} is modified to include $\E(\tilde{\bm{\epsilon}}^{(h)} \tilde{\bm{\epsilon}}^{(k)*}{}')\times\OP(n^{-1/2})$. As a result, $\oP(n^{-3/2})$ changes to $\OP(n^{-3/2})$ in the evaluation of \eqref{Veval}. Despite this change, Theorem \ref{th3} still holds.

\subsection{Case of doubly robust estimation}
\label{sec6_4}
Doubly robust estimation is often used as a potentially more efficient alternative to inverse-probability-weighted estimation (\citealt{RobRot95}). In this estimation, we directly model the regression of the outcome variable on the confounding variable, i.e., $\mu^{(h)}(\bm{X})+f(\bm{X})$, and we write its regression function as $g^{(h)}(\bm{X})$. If either the modeling or the propensity score modeling is correct, this method can give a consistent estimator of the causal effect, which is why it is called doubly robust (\citealt{SchRR99}). Note that unlike in the previous section, in doubly robust estimation, it is not necessarily true that $\hat{e}^{(h)}(\bm{X}_i)-e^{(h)}(\bm{X}_i)=\OP(n^{-1/2})$.

For simplicity, we will use $\{\bm{A}(\bm{M},\bm{s})\hat{\tilde{\bm{W}}}(\tilde{\bm{T}},\tilde{\bm{X}})\tilde{\bm{Y}}\le\bm{b}(\bm{M},\bm{s})\}$ for the conditioned event. Note that, although we will not consider it here, there is no difficulty in using double robust estimating equation for this model selection when the estimator of the function $g^{(h)}(\bm{X})$, written as $\hat{g}^{(h)}(\bm{X})$, is a linear function of $\bm{Y}$. The doubly robust estimator of $\beta_j$ is given by
\begin{align}
\bm{e}_j' ( \tilde{\bm{X}}_{\bm{M}}' \tilde{\bm{X}}_{\bm{M}} )^{-1} \tilde{\bm{X}}_{\bm{M}}' \sum_{h=1}^H c^{(h)} [ \hat{\tilde{\bm{W}}}^{(h)} ( \tilde{\bm{T}}^{(h)}, \tilde{\bm{X}} ) \tilde{\bm{Y}}^{(h)} + \{ \bm{I}_n-\hat{\tilde{\bm{W}}}^{(h)} ( \tilde{\bm{T}}^{(h)}, \tilde{\bm{X}} ) \} \hat{\tilde{\bm{g}}}^{(h)} ( \tilde{\bm{X}} ) ].
\label{dre}
\end{align}
Here, $\hat{\tilde{\bm{g}}}^{(h)}(\tilde{\bm{X}})=(\hat{g}^{(h)}(\bm{X}_1),\ldots,\hat{g}^{(h)}(\bm{X}_n))'$, and we assume that when this modeling is correct, the differences between it and $\tilde{\bm{\mu}}^{(h)}(\tilde{\bm{X}})+\tilde{\bm{f}}(\tilde{\bm{X}})$ and between it and $\E\{\hat{\tilde{\bm{g}}}^{(h)}(\tilde{\bm{X}})\mid\tilde{\bm{X}}\}$ are each $\OP(n^{-1/2})$. Then, by defining
\begin{align*}
\tau_j^{\bm{M}} ( \tilde{\bm{Y}}^{\dagger},\tilde{\bm{T}},\tilde{\bm{X}} ) \equiv \bm{e}_j' ( \tilde{\bm{X}}_{\bm{M}}' \tilde{\bm{X}}_{\bm{M}} )^{-1} \tilde{\bm{X}}_{\bm{M}}' \sum_{h=1}^H c^{(h)} \{ \hat{\tilde{\bm{W}}}^{(h)} ( \tilde{\bm{T}}^{(h)},\tilde{\bm{X}} ) - \bm{I}_n \} [ \tilde{\bm{Y}}^{(h)\dagger} - \E \{ \hat{\tilde{\bm{g}}}^{(h)} ( \tilde{\bm{X}} ) \mid \tilde{\bm{X}} \} ],
\end{align*}
and replacing the estimator $\tilde{\bm{\eta}}_j' \tilde{\bm{W}}( \tilde{\bm{T}},\tilde{\bm{X}}) \tilde{\bm{Y}}$ with \eqref{dre}, we can see that Lemma \ref{th2} holds.

Next, as $\tilde{\bm{Y}}^{(h)\dagger}$ above, we consider using its natural estimator like $\tilde{\bm{Y}}^{(h)*}$ in Section \ref{sec3}. However, $\tilde{\bm{Y}}^{(h)*}$ converges slowly, so if the modeling of the propensity score is incorrect and the expectation of $\hat{\tilde{\bm{W}}}^{(h)}(\tilde{\bm{T}}^{(h)},\tilde{\bm{X}})-\bm{I}_n$ does not converge to the zero matrix, we cannot expect quick convergence in evaluations like those of \eqref{Eeval} and \eqref{Veval}. As a naive solution, we propose to use a goodness-of-fit statistic for $\hat{e}^{(h)}(\bm{X})$ and substitute $\tilde{\bm{Y}}^{(h)*}-\hat{\tilde{\bm{g}}}^{(h)}(\tilde{\bm{X}})$ or $\bm{0}_n$ for $\tilde{\bm{Y}}^{(h)\dagger}-\E\{\hat{\tilde{\bm{g}}}^{(h)}(\tilde{\bm{X}})\mid\tilde{\bm{X}}\}$ by case separation. For example, suppose that the possible values of $\hat{e}^{(h)}(\bm{X})$ estimated using a $q$-dimensional discriminant model are discrete, and the estimate is $\hat{e}^{(h)}_r$ when $\bm{X}\in{\cal X}_r\ (\subset\mathbb{R}^p)$, where $\bigcup_{r=1}^R{\cal X}_r=\mathbb{R}^p$. Then, by letting $\hat{p}^{(h)}_r\equiv\sum_{i:\bm{X}_i\in{\cal X}_r}T_i^{(h)}/|\{i:\bm{X}_i\in{\cal X}_r\}|$, we only have to consider $d(\tilde{\bm{T}},\tilde{\bm{X}})\equiv-2\sum_{r=1}^R\sum_{i:\bm{X}_i\in{\cal X}_r}\sum_{h=1}^HT_i^{(h)}\log(\hat{p}^{(h)}_r/\hat{e}^{(h)}_r)+(LH-q)\log n$, which is equivalent to a difference between Bayesian information criteria (BIC), as the goodness-of-fit statistic. If this goodness-of-fit statistic is positive, the propensity score model is regarded as valid and we use $\tilde{\bm{Y}}^{(h)*}-\hat{\tilde{\bm{g}}}^{(h)}(\tilde{\bm{X}})$. If it is negative, the propensity score model is regarded as not valid and we use $\bm{0}_n$. Then, from the consistency of BIC, we can ensure the convergence as in Section \ref{sec3}. Specifically, we define
\begin{align*}
& \tau_j^{\bm{M}} ( \tilde{\bm{Y}}^{*},\tilde{\bm{T}},\tilde{\bm{X}} ) 
\\
& \equiv \bm{e}_j' ( \tilde{\bm{X}}_{\bm{M}}' \tilde{\bm{X}}_{\bm{M}} )^{-1} \tilde{\bm{X}}_{\bm{M}}' I\{d(\tilde{\bm{T}},\tilde{\bm{X}})>0\} \sum_{h=1}^H c^{(h)} \{ \hat{\tilde{\bm{W}}}^{(h)} ( \tilde{\bm{T}}^{(h)},\tilde{\bm{X}} ) - \bm{I}_n \} \{ \tilde{\bm{Y}}^{(h)*} - \hat{\tilde{\bm{g}}}^{(h)} ( \tilde{\bm{X}} ) \}.
\end{align*}
In line with this, we will let $\tilde{\bm{K}}^{(h)}(\tilde{\bm{X}})\equiv{\rm diag}[\{1/e^{(h)}(\bm{X}_1)-1\}/|{\mathcal N}_1^{(h)}|,\ldots,\{1/e^{(h)}(\bm{X}_n)-1\}/|{\mathcal N}_n^{(h)}|]$. Moreover, we define
\begin{align*}
\rho_j^{\bm{M}} ( \tilde{\bm{T}},\tilde{\bm{X}} ) \equiv \zeta_j^{\bm{M}} ( \tilde{\bm{T}},\tilde{\bm{X}} ) + \sigma^2 I\{d(\tilde{\bm{T}},\tilde{\bm{X}})>0\} \sum_{h=1}^H c^{(h)2} \tilde{\bm{\eta}}_j' \tilde{\bm{K}}^{(h)}(\tilde{\bm{X}}) \tilde{\bm{\eta}}_j,
\end{align*}
where $I\{\cdot\}$ is the indicator function. Using these definitions, it can be seen that Theorem \ref{th3} holds.

\subsection{Case of time-varying confounding}
\label{sec6_5}
Dealing with time-varying confounding is a fundamental topic in causal inference (\citealt{BanRob05}; \citealt{DanCSKS13}). Here, suppose that there are $m$ time points, indexed by $j\ (\in\{1,\ldots,m\})$. Let $T_j$ be the assignment variable representing the binary time-varying treatment and $\bm{X}_j$ be the time-varying confounding variable at time $j$. The collections $\bm{T}=(T_1,\ldots,T_m)'$ and $\bm{X}=(\bm{X}_1',\ldots,\bm{X}_m')'$ are called the treatment history and the confounding history, respectively. Let us denote the realization of $\bm{T}$ as $\bm{t}=(t_1,\ldots,t_m)'$ and the potential outcome variable for $\bm{t}$ as $Y^{(\bm{t})}$. We assume the following structure,
\begin{align*}
Y=\sum_{\bm{t}\in\{0,1\}^m}I(\bm{T}=\bm{t})Y^{(\bm{t})}=\sum_{\bm{t}\in\{0,1\}^m}I(\bm{T}=\bm{t})\{\mu^{(\bm{t})}+f(\bm{X})+\epsilon\}
\end{align*}
as in \cite{DanCSKS13}. Then, for this $\mu^{(\bm{t})}$, we consider the model $(\bm{t}-\bm{1}_m/2)'\bm{\beta}$. Note that $\beta_j$ is a parameter that represents the treatment effect at time $j$. As in the preceding sections, the variables for the $i$-th sample are denoted by the subscript ${}_i$ and their collection is denoted by the tilde symbol $\tilde{\ }$. In this setting, the ignorable treatment assignment condition defined by \eqref{assump1} does not give a consistent estimation, so 
\begin{align*}
Y_{i}^{(\bm{t})} \indep T_{j,i} \mid \bm{X}_{1,i},\ldots,\bm{X}_{j,i},T_{1,i},\ldots,T_{j-1,i}\qquad(\bm{t}\in\{0,1\}^m;\ i=1,\ldots,n;\ j=1,\ldots,m)
\end{align*}
is assumed instead. Here, when $j=1$, the assignment variable is assumed to be unconditioned.

The propensity score in this setting is $e^{(\bm{t})}(\bm{X})=\P(T_{1}=t_{1}\mid\bm{X}_{1})\prod_{j=2}^{m}\P(T_{j}=t_{j}\mid\bm{X}_{1},\ldots,\bm{X}_{j},T_{1}=t_1,\ldots,T_{j-1}=t_{j-1})$, and the weight matrix is $\tilde{\bm{W}}^{(\bm{t})}(\tilde{\bm{T}},\tilde{\bm{X}})={\rm diag}\{I(\bm{T}_1=\bm{t})/e^{(\bm{t})}(\bm{X}_1),\ldots,I(\bm{T}_n\allowbreak=\bm{t})/e^{(\bm{t})}(\bm{X}_n)\}$. We conduct model selection by minimizing $\sum_{\bm{t}\in\{0,1\}^m} \allowbreak \| \tilde{\bm{W}}^{(\bm{t})}(\tilde{\bm{T}},\tilde{\bm{X}}) \tilde{\bm{Y}} - \bm{1}_n (\bm{t}-\bm{1}_m/2)' \bm{\beta} \|_2^2/2 + \lambda \| \bm{\beta} \|_1$. Once a model $\bm{M}$ is selected, we use the inverse-probability-weighted estimator,
\begin{align*}
\frac{1}{n2^{|\bm{M}|-2}}\bm{e}_j'\sum_{\bm{t}\in\{0,1\}^m}\bigg(\bm{t}^{\bm{M}}-\frac{\bm{1}_{|\bm{M}|}}{2}\bigg)\bm{1}_n'\tilde{\bm{W}}^{(\bm{t})}(\tilde{\bm{T}},\tilde{\bm{X}})\tilde{\bm{Y}},
\end{align*}
as a consistent estimator of $\beta_j^{\bm{M}}$. Although the model selection and estimation have different expressions from those in Section \ref{sec3}, it makes no difference that the squared losses are based on a linear function of $\tilde{\bm{Y}}$ minus a linear function of $\bm{\beta}$. Therefore, for example, an event in which model $\bm{M}$ with sign $\bm{s}$ is selected is expressed as $\bm{A}(\bm{M},\bm{s},\tilde{\bm{T}},\tilde{\bm{X}}) \tilde{\bm{Y}} \le \bm{b}(\bm{M},\bm{s})$, and ${\mathcal V}_{\bm{s},j}^-$ and ${\mathcal V}_{\bm{s},j}^+$ are easily determined, i.e., the argument can be developed in the same way as in Section \ref{sec3}. Specifically, by letting $\tilde{\bm{K}}^{(\bm{t})}(\tilde{\bm{X}})\equiv{\rm diag}[\{1/e^{(\bm{t})}(\bm{X}_1)-1\}/|{\cal N}_1^{(\bm{t})}|,\ldots,\{1/e^{(\bm{t})}(\bm{X}_n)-1\}/|{\cal N}_n^{(\bm{t})}|]$, 
\begin{align*}
& \kappa_j^{\bm{M}} ( \tilde{\bm{T}},\tilde{\bm{X}} ) = \frac{1}{n2^{|\bm{M}|-2}}\bm{e}_j' \sum_{\bm{t}\in\{0,1\}^m}\bigg(\bm{t}^{\bm{M}}-\frac{\bm{1}_{|\bm{M}|}}{2}\bigg) \bm{1}_n' \tilde{\bm{W}}^{(\bm{t})}(\tilde{\bm{T}},\tilde{\bm{X}})\{\tilde{\bm{\mu}}^{(\bm{t})}+\tilde{\bm{f}}(\bm{X})\},
\\
& \zeta_j^{\bm{M}} ( \tilde{\bm{T}},\tilde{\bm{X}} ) = \frac{\sigma^2}{n^22^{2|\bm{M}|-4}} \sum_{\bm{t}\in\{0,1\}^m} \bm{e}_j' \bigg(\bm{t}^{\bm{M}}-\frac{\bm{1}_{|\bm{M}|}}{2}\bigg) \bm{1}_n' \tilde{\bm{W}}^{(\bm{t})}(\tilde{\bm{T}},\tilde{\bm{X}})^2 \bm{1}_n \bigg(\bm{t}^{\bm{M}}-\frac{\bm{1}_{|\bm{M}|}}{2}\bigg)' \bm{e}_j,
\\
& \tau_j^{\bm{M}} ( \tilde{\bm{Y}}^{\dagger},\tilde{\bm{T}},\tilde{\bm{X}} ) = \frac{1}{n2^{|\bm{M}|-2}}\bm{e}_j' \sum_{\bm{t}\in\{0,1\}^m}\bigg(\bm{t}^{\bm{M}}-\frac{\bm{1}_{|\bm{M}|}}{2}\bigg) \bm{1}_n' \{\tilde{\bm{W}}^{(\bm{t})}(\tilde{\bm{T}},\tilde{\bm{X}})-\bm{I}_n\} \tilde{\bm{Y}}^{(\bm{t})\dagger}
\end{align*}
and
\begin{align*}
\rho_j^{\bm{M}} ( \tilde{\bm{T}},\tilde{\bm{X}} ) = \zeta_j^{\bm{M}}(\tilde{\bm{T}},\tilde{\bm{X}}) + \frac{\sigma^2}{n^22^{2|\bm{M}|-4}} \sum_{\bm{t}\in\{0,1\}^m} \bm{e}_j' \bigg(\bm{t}^{\bm{M}}-\frac{\bm{1}_{|\bm{M}|}}{2}\bigg) \bm{1}_n' \tilde{\bm{K}}^{(\bm{t})}(\tilde{\bm{X}}) \bm{1}_n \bigg(\bm{t}^{\bm{M}}-\frac{\bm{1}_{|\bm{M}|}}{2}\bigg)' \bm{e}_j,
\end{align*}
it can be seen that Theorem \ref{th1}, Lemma \ref{th2} and Theorem \ref{th3} hold.

\section{Conclusion}
\label{sec7}
We developed selective inference in a basic causal inference model where the least-squares method is inappropriate for when inverse-probability-weighted estimation is used as the standard for propensity score analysis. Conditioning on the assignment variables first is usually not done in propensity score analysis, but doing so, we were able to construct a pivot statistic that follows a uniform distribution. Moreover, by dealing with higher-order asymptotics, we gave asymptotically guaranteed confidence intervals in the context of selective inference. Numerical experiments showed that a method that ignores the quiet scandal of statistics results in significant deviations from the preset coverage of the confidence intervals, whereas our method maintains the coverage. By comparing the two methods on benchmark data, we showed that there is a significant difference between them.

To carry on the beautiful methodology described by \cite{LeeSST16}, we considered in this paper model selection with basic sparse estimation in a setting that follows a normal distribution if there is no assignment. However, it is also important to consider model selection with an information criterion in a setting that includes a non-normal distribution. The information criterion for this case was recently developed by \cite{BabKN17}, and it is expected that there will be no difficulty in implementing it in the method of \cite{ChaC18}. On the other hand, in order to treat the error distribution non-parametrically and to construct an elaborate theory based on the uniform convergence of pivot statistics, it is necessary to develop the works of \cite{TiaT17} and \cite{TibRTW18}.

The selective inference is developing rapidly, and there are ongoing discussions on increasing the power of test.
In fact, \cite{KivL20} has shown that the standard method of selective inference can give too wide confidence intervals due to $\E(U_j^{\hat{\bm{M}}}-L_j^{\hat{\bm{M}}})=\infty$.
Then, for example, \cite{TiaT18} proposed a modern version of sample splitting, specifically adding an appropriate randomization to the data, which is further discussed in \cite{RasY21}.
Depending on how the randomization is realized, these methods may yield different results, and so they cannot be accepted in all the fields of application, however, a considerable improvement in the power has been observed.
In addition, \cite{LiuMT18} pointed out that it is over-conditioning to condition on $\hat{\bm{M}}=\bm{M}$ when inferring each $\beta_j$, and increased the power by conditioning on a wider region.
Since the selective inference in propensity score analysis involves many non-trivial arguments in constructing valid inferences themselves, and since the above-mentioned improvements are expected to be realized by simply combining them, we have not dealt with them in this paper.
However, they are significant and will be an important subject in the future.

\section*{Acknowledgement}
Yoshiyuki Ninomiya was supported by JSPS KAKENHI (16K00050), Yuta Umezu was supported by JSPS KAKENHI (18K18010), and Ichiro Takeuchi was supported by JSPS KAKENHI (17H00758, 20H00601), JST CREST (JPMJCR1502) and RIKEN Center for Advanced Intelligence Project.

\bibliographystyle{jecon}
\bibliography{List}

\end{document}